\documentclass[twocolumn]{aastex631_arxiv}
\usepackage{amsmath}
\usepackage{multirow}
\usepackage{mathtools}
\usepackage[thinc]{esdiff}

\newcommand{\bz}{$\langle B_z \rangle$}

\received{----}
\revised{----}
\accepted{----}
\submitjournal{ApJ}

\begin{document}

\title{Discoveries of fine structures and secondary pulses in coherent radio emission from a magnetic massive star}

\shorttitle{MeerKAT observation of HD 142990}
\shortauthors{Das et al.}

\correspondingauthor{Barnali Das}
\email{Barnali.Das@csiro.au}

\author[0000-0001-8704-1822]{Barnali Das}
\affil{CSIRO, Space and Astronomy, PO Box 1130, Bentley WA 6102, Australia}

\author[0000-0002-0844-6563]{Poonam Chandra}
\affil{National Radio Astronomy Observatory, 520 Edgemont Road, Charlottesville VA 22903, USA}
\affil{National Centre for Radio Astrophysics, Tata Institute of Fundamental Research, Pune University Campus, Pune-411007, India}

\author[0000-0001-7363-6489]{William Cotton}
\affil{National Radio Astronomy Observatory, 520 Edgemont Road, Charlottesville VA 22903, USA}

\author[0000-0002-5633-7548]{V\'eronique Petit}
\affiliation{Department of Physics and Astronomy, Bartol Research Institute, University of Delaware, 104 The Green, Newark, DE 19716, USA}







\begin{abstract}
In this paper, we report auroral radio emission from a magnetic B star HD 142990 using the MeerKAT radio telescope at 900--1670 MHz. The star is known to produce such emission (observed as periodic radio pulses) via electron cyclotron maser emission (ECME). However, past studies on ECME from this star were confined to observations at specific rotational phase ranges where one expects to see such pulses. We, for the first time, observed the star for its one complete rotation cycle and discovered that the star also produces `off-pulse' emission, which we term as secondary enhancements. Two such enhancements were observed, one of which is left circularly polarized (LCP) and the other is right circularly polarized (RCP), the latter is confirmed to be persistent. Using simulation, we infer that such pulses are likely related to the large misalignment between the stellar rotation and magnetic dipole axes ($>80^\circ$), leading to the formation of highly complex magnetospheric plasma distribution. In addition, by extracting dynamic spectra for the primary pulses, we discovered prominent fine structures in one of the LCP pulses, with timescales as small as the instrumental time resolution (8 seconds). This is the first time that such structures are seen from a magnetic hot star, and has the potential to reveal detailed information about how the emission is driven, and the nature of the elementary sources of radiation. To pinpoint the origin of these fine structures and their significance, higher time and spectral resolution observations should be conducted in the future.
\end{abstract}


\section{Introduction} \label{sec:intro}
A subset of the magnetic hot (spectral types OBA) stars, which are characterized by the presence of highly stable and typically dipolar surface magnetic fields, are known to produce periodic radio pulses by the electron cyclotron maser emission (ECME) mechanism \citep[e.g.][]{trigilio2000,das2022}.
The emission is believed to originate near the auroral regions (regions close to the magnetic poles) and hence the emission is also refereed to as auroral radio emission. Magnetic hot stars producing this kind of emission are now referred to as `Main-sequence Radio Pulse emitters' \citep[MRPs,][]{das2021}. The underlying emission mechanism, ECME, is a coherent mechanism that requires an unstable electron distribution and a magnetised plasma with the plasma frequency that is much smaller than the electron gyrofrequency \citep[e.g.][]{treumann2006}. It is an intrinsically narrowband phenomenon and the frequency of emission is proportional to the magnetic field strength at the emission sites.
As a result, high frequencies are produced closer to the star (at regions of stronger magnetic field strength) and vice-versa.
The emission is also beamed. For magnetic hot stars, the beaming geometry is described by the `tangent plane beaming model' \citep{trigilio2011}. This model predicts observation of the emission as periodic radio pulses around the rotational phases at which the stellar longitudinal magnetic field $B_\mathrm{z}$ is zero (called magnetic nulls). Indeed for the first few MRPs (that were observed over full rotation cycles), this was found to be valid over the respective frequencies of observations at the time \citep[CU\,Vir, HD\,133880, HD\,142301 and $\rho$ Oph A;][]{trigilio2000,das2018,leto2019,leto2020}.

The biggest deviation from the prediction of the tangent plane beaming model was discovered when CU\,Vir \citep[the first discovered MRP,][]{trigilio2000} was observed below 1 GHz for the first time \citep{das2021}. Note that this star had been observed extensively above 1 GHz covering its full rotation cycle and it was always observed to produce two narrow pulses per cycle. However, at 400 MHz, clear secondary enhancements were observed that do not have higher frequency counterparts.
Such secondary pulses were also reported for the star HD\,12447 by \citet{das2022}, the only star in their sample that was observed for one full rotation cycle.

Interestingly, prior to these observational evidence, existence of such secondary pulses was suggested by \citet{das2020a} for magnetic hot stars with large misalignments between their magnetic and rotational axes.
In case of an aligned dipole, most of the magnetospheric plasma remains confined to the common magneto-rotational equator \citep{townsend2005} so that ECME, that is produced above the magnetic poles, does not have to pass through the high density plasma. The situation changes dramatically as the misalignment approaches $90^\circ$. In this case, plasma is distributed in a complex manner and ECME radiations at different frequencies can experience very different plasma densities on their way to the observer; the resulting refraction/reflection splits and shifts the pulses in the light curve. To investigate these effects, \citet{das2020a} provided a three dimensional framework to simulate ECME lightcurves for any kind of plasma distribution in the stellar magnetosphere. Note that the framework assumes the tangent plane beaming model. Using the framework, they showed that secondary pulses can appear in the lightcurve at certain frequencies depending on the plasma distribution as well as the magnetic field strength.

With an aim to explore the possibility of ECME visible at phases away from those of the primary pulses, we conducted full rotation cycle observation of a known MRP HD\,142990 with the MeerKAT radio telescope \citep{jonas2016}. In this paper, we report the results of our observation over 900--1670 MHz.
Our target has a strong surface magnetic field with a polar strength of 4.7 kG with significant non-dipolar component \citep{shultz2018,shultz2019c}. It is one of the extreme oblique rotators with an obliquity (angle between the magnetic dipole and the rotation axes) of $\gtrsim 80^\circ$ \citep{shultz2019c}. The star is already known to exhibit peculiar ECME properties over 0.4--3 GHz \citep{das2019a,das2023}.
It is the only MRP for which the arrival sequence of oppositely circularly polarized pulses reverses with frequency. The primary pulses from this star show multiple components resulting into complex pulse profiles. Note that \citet{das2023} referred the weaker pulse-components as secondary pulses. 
In this paper, however, we will treat the enhancements observed close to the magnetic nulls (i.e. all the enhancements reported in the past) as constituents of the primary pulses.
The peculiar properties of the primary pulses of the star motivated us to conduct an observation covering its full rotation cycle in search of any secondary enhancements predicted by the propagation effect scenario of \citet{das2020a}.

This paper is structured as follows. In the next section, we describe our observation and data analysis (\S\ref{sec:data}). This is followed by our results (\S\ref{sec:results}) and discussion (\S\ref{sec:discussion}). We summarize the paper in \S\ref{sec:summary}.

\section{Observation and data analysis}\label{sec:data}
We observed HD\,142990 using the MeerKAT telescope in subarray mode with half of the antennas in the UHF band (580--1015 MHz) and the other half in L-band (900--1670 MHz). 
In this paper, we report the results obtained from the L-band data (the UHF data will be reported in a follow-up publication). The full rotation cycle was covered by observing on three days, covering $\approx 0.36$ rotation cycles per day. MeerKAT provides data in linear basis and polarization calibration is required to obtain Stokes V information. To do that, we included at least one scan of 3C286 (for the calibration of the X-Y phase difference needed to separate linear from circular polarization) in addition to observing J1939--6342 (to calibrate the spurious polarized response to Stokes I, commonly known as leakages). The latter also acts as a flux/bandpass calibrator. 
Observations recorded all four combinations of the orthogonal linearly polarized feeds, had 8 second integrations and 4096 channels across the band pass.
Observations consisted of 30 minute scans on HD\,142990 interleaved with calibration scans.
The project code was SCI-20220822-BD-01.
The details of our observation are given in Table \ref{tab:targets_obs}.

\begin{deluxetable*}{crccc}
\centering
\tablecaption{Observation logs for HD\,142990. HJD stands for Heliocentric Julian Day. The three days of observations will be referred as Day 1, Day 2 and Day 3 as indicated in the first column. The rotational phases are calculated using the ephemeris of \citet{shultz2019_0}.\label{tab:targets_obs}}  
\tablehead{
Date & HJD range & Rotational &  \multicolumn{2}{c}{Calibrator}\\
of Obs. & $-2.45\times 10^6$ & phase range  & Flux/bandpass & Phase 
}
\startdata
\hline
2023--01--29 (Day 1) & $9973.78\pm 0.17$ & 0.47--0.82 & J0408--6545, J1939--6342 & J1517--2422\\
2023--03--08 (Day 2) & $10012.60\pm 0.18$ & 0.12--0.49 & J0408--6545, J1939--6342 & J1517--2422 \\
2023--04--08 (Day 3) & $10043.61\pm 0.18$& 0.80--1.17 & J1939--6342 & J1517--2422, J1605-1734 \\
\enddata
\end{deluxetable*}

Data calibration and editing followed the general scheme outlined in
\cite{DEEP2, XGalaxy} and used the Obit
package\footnote{\url{http://www.cv.nrao.edu/~bcotton/Obit.html}}
\citep{OBIT}.
For calibration purposes the outer channels were trimmed from the spectra and the data divided into 8 equal spectral windows.
The noise diode calibration at the beginning of each observing session was used to align the phases of the two systems of parallel linear feeds.
Data in spectral regions with known, persistent interfering signals were flagged as well as times and frequencies identified in subsequent editing and calibration steps.
The flux density scale was set by the \cite{Reynolds94} spectrum of J1939--6342:
$$
  \log(S) = -30.7667 + 26.4908 \log\bigl(\nu\bigr)
  - 7.0977 \log\bigl(\nu\bigr)^2 $$
  $$+0.605334 \log\bigl(\nu\bigr)^3,$$
where $S$ is the flux density (Jy) and $\nu$ is the frequency (MHz).

The feeds were modeled in terms of the ellipticity and orientation of
the radiation to which they are sensitive.
In a sample of solutions, the average absolute deviation of the
solution from nominal was 0.0005 in ellipticity and 0.6$^\circ$ in
orientation. 
In the small error approximation, this corresponds to a ``leakage''
amplitude of 0.011.

The calibrated data had 8 second integrations and 
after Hanning smoothing, had 1896$\times$418 kHz channels,
although many were lost due to RFI flagging.
After calibration, the sky model (CLEAN component table) produced by the SARAO calibration pipeline imaging, minus components near HD\,142990, was subtracted from the data.
Application of the polarization calibration converted the data from linear to a circular basis (RR,LL).

The field source subtracted visibilities were then imported to \textsc{casa} \citep{mcmullin2007} for extracting lightcurves and dynamic spectra.
The full-band averaged lightcurves in right and left circular polarisations (RCP and LCP respectively) were obtained via imaging using the task \texttt{tclean}. 
The flux densities were estimated using the task \texttt{imfit}. The error bars in the flux densities include both the fitting error and image rms (added in quadrature). 

To extract the dynamic spectra, the real parts of the visibilities were averaged over all baselines at the original time and frequency resolutions (8 seconds and 418 kHz respectively). 
This part of the analysis was done using \texttt{python} with the help of the modular form of the \textsc{casa} package\footnote{\url{https://casadocs.readthedocs.io/en/stable/examples/community/casa6_demo.html}}.

We phased the data using the ephemeris of \citet{shultz2019_0}. The rotational phases are given by the fractional part of:
\begin{align}
    N &=\frac{2\Delta t}{2P_0+\dot{P}\Delta t},\label{eq:ephemeris}
\end{align}
where $\Delta t=\mathrm{HJD}-\mathrm{HJD_0}$, the reference epoch $\mathrm{HJD_0}=2442820.93(3)$ days, 
the rate of change of rotation period $\dot{P}=-0.58\pm0.02\,\mathrm{s/year}$ and the rotation period at the reference epoch $P_0=0.979110(4)$ days. 
The phase zero corresponds to the negative extremum of the stellar longitudinal magnetic field \bz~\citep{shultz2019_0}. The magnetic nulls occur at phases 0.250 (\bz~changes from negative to positive) and 0.664 (\bz~changes from positive to negative). Following the convention of \citet{das2019a}, we will refer the former (phase 0.250) as null 1 and the latter (phase 0.664) as null 2.

Note that for the uncertainties in various quantities in the ephemeris, the rotational phases at our epochs of observations have an uncertainty of $\approx 0.13$.

We use the IAU/IEEE convention for defining RCP and LCP. Under this convention, Stokes $V$ is defined as (RCP--LCP)/2.

\section{Results}\label{sec:results}

\begin{figure}
    \centering
    \includegraphics[width=0.49\textwidth]{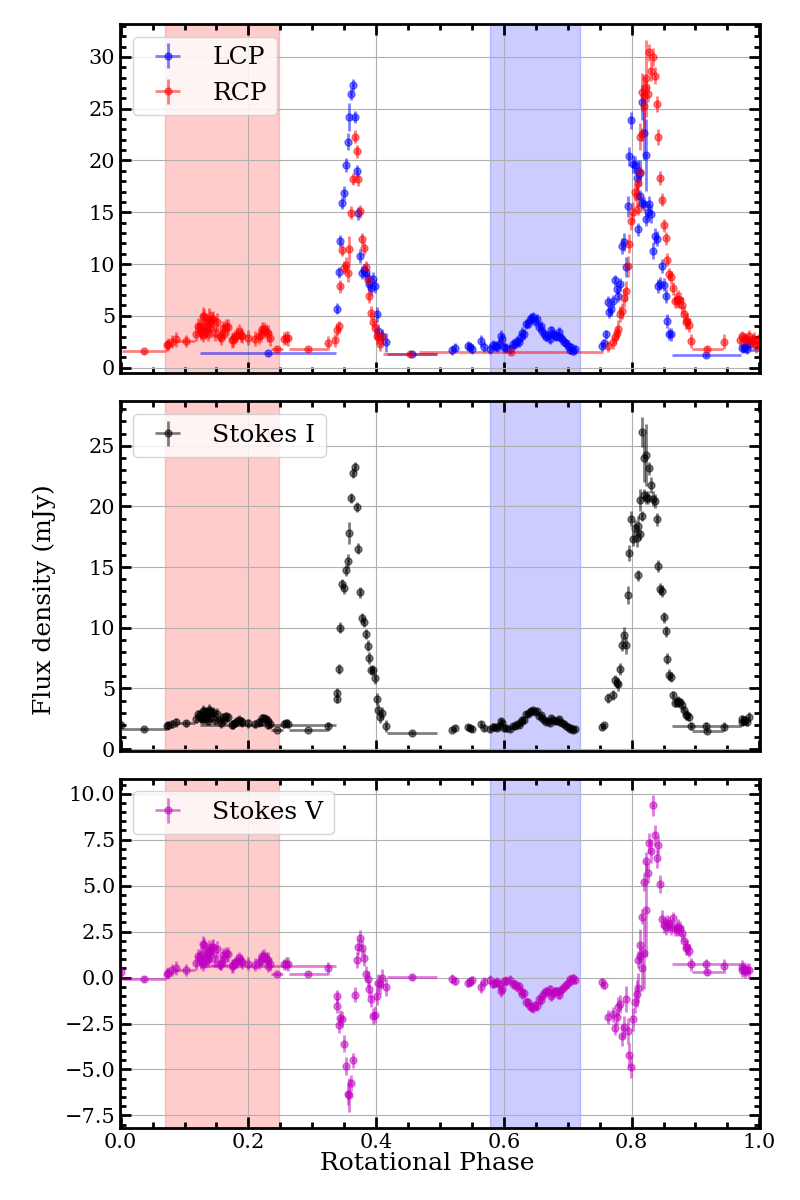}
    \caption{The lightcurves of HD\,142990 averaged over 900--1700 MHz. The top panel shows the rotational phase variation in right (red) and left (blue) circular polarizations. The middle and bottom panels show the corresponding Stokes I and Stokes V lightcurves respectively.}
    \label{fig:Lband_LC}
\end{figure}

\begin{figure*}
    \centering
    \includegraphics[width=0.325\textwidth]{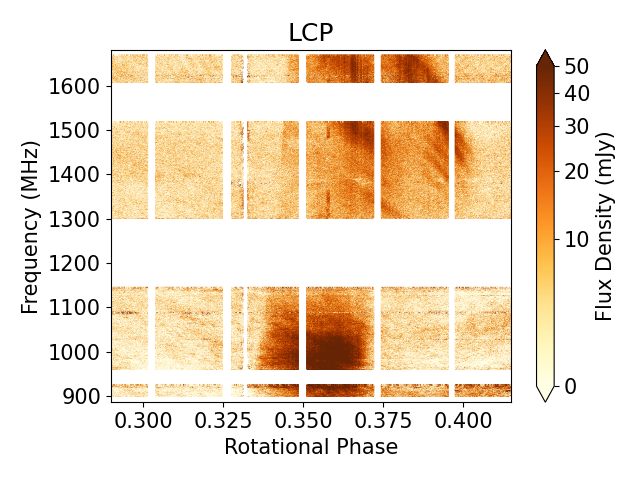}
    \includegraphics[width=0.325\textwidth]{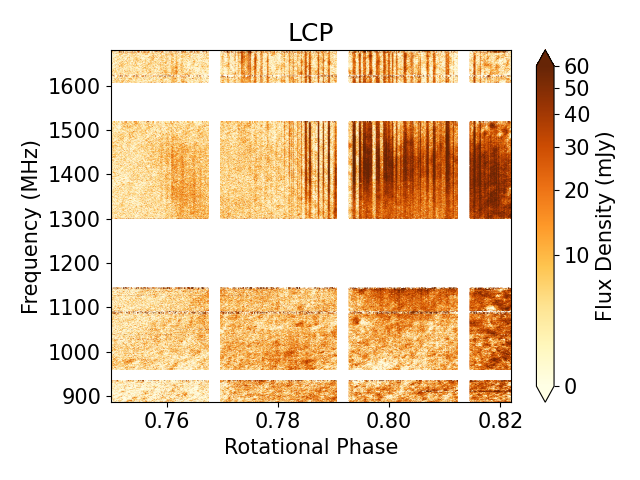}
    \includegraphics[width=0.325\textwidth]{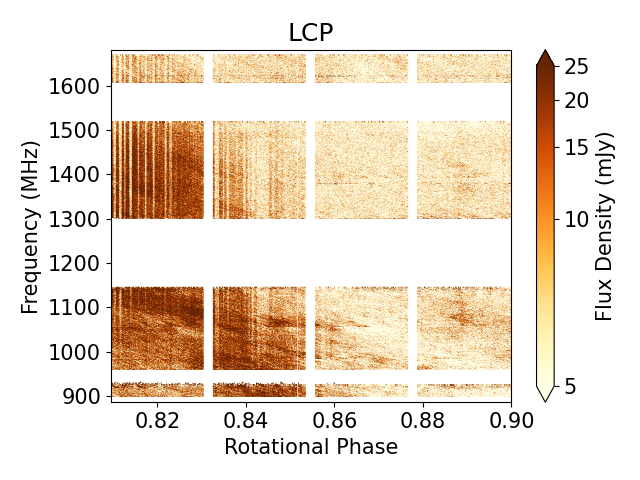}
    \includegraphics[width=0.325\textwidth]{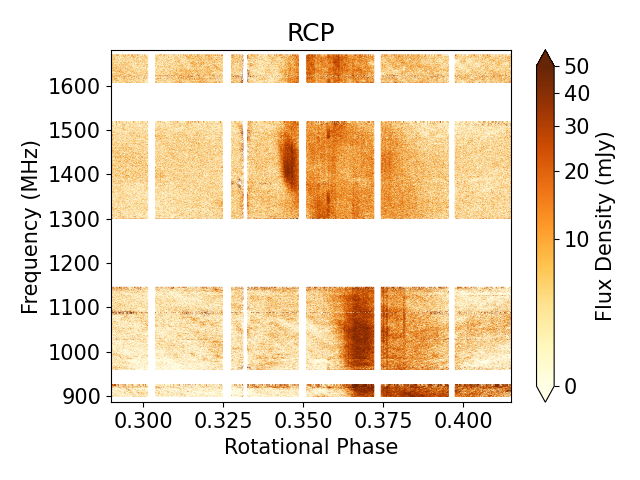}
    \includegraphics[width=0.325\textwidth]{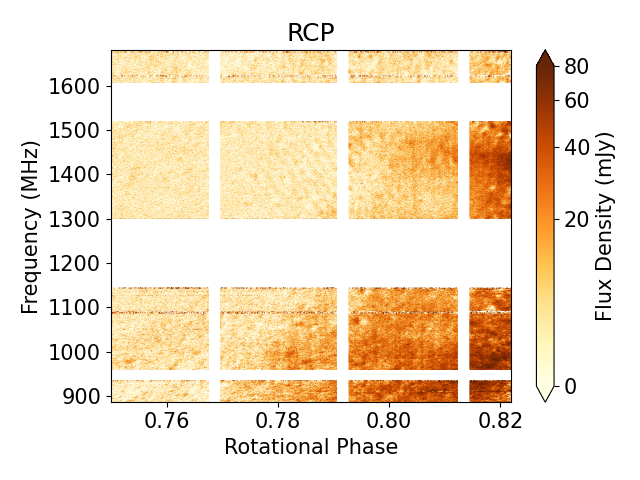}
    \includegraphics[width=0.325\textwidth]{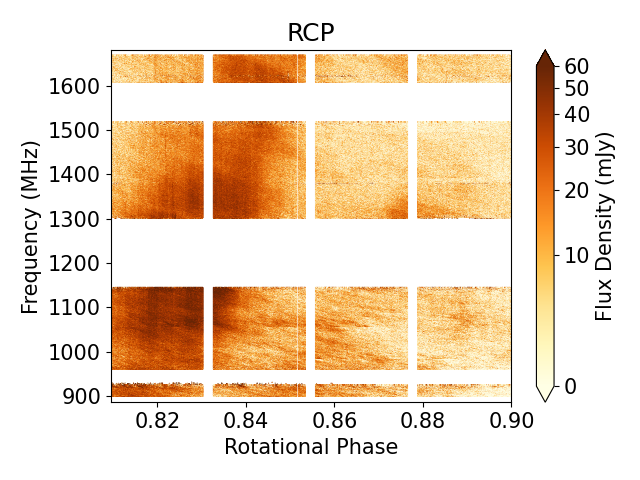}
    \includegraphics[width=0.325\textwidth]{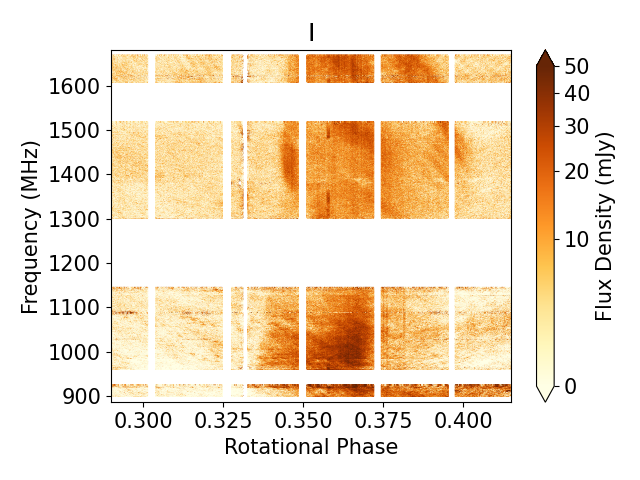}
    \includegraphics[width=0.325\textwidth]{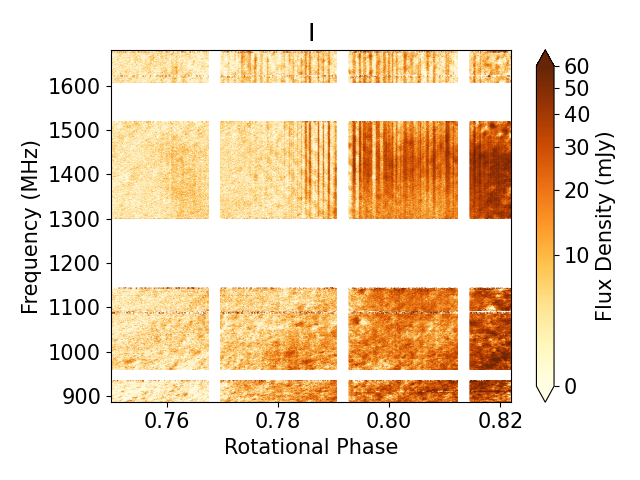}
    \includegraphics[width=0.325\textwidth]{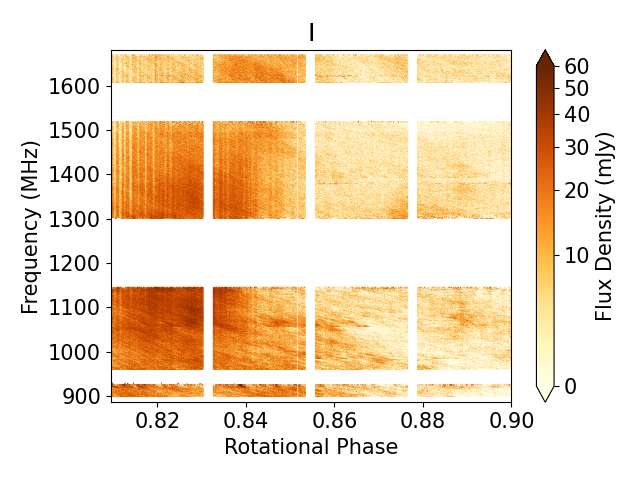}
    \includegraphics[width=0.325\textwidth]{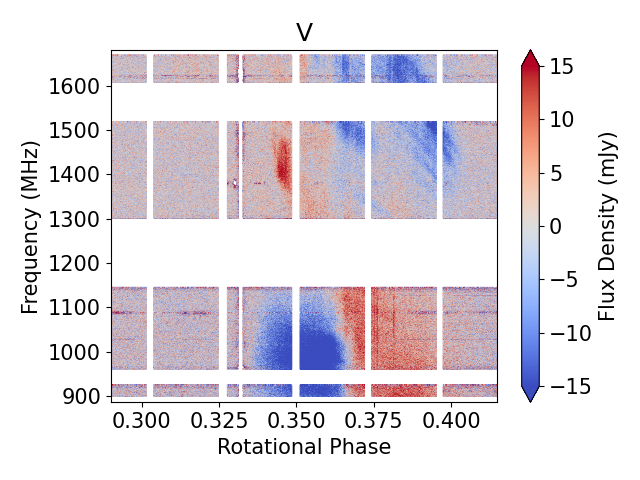}
    \includegraphics[width=0.325\textwidth]{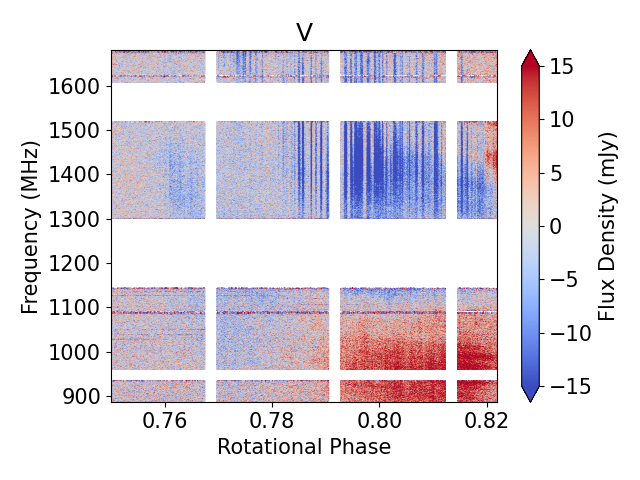}
    \includegraphics[width=0.325\textwidth]{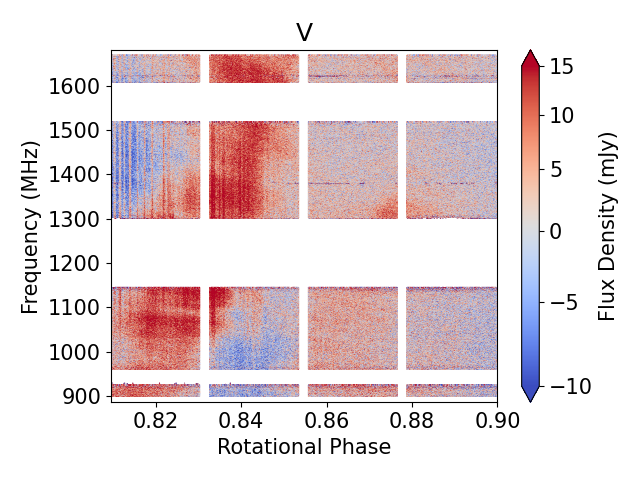}
    \includegraphics[width=0.325\textwidth]{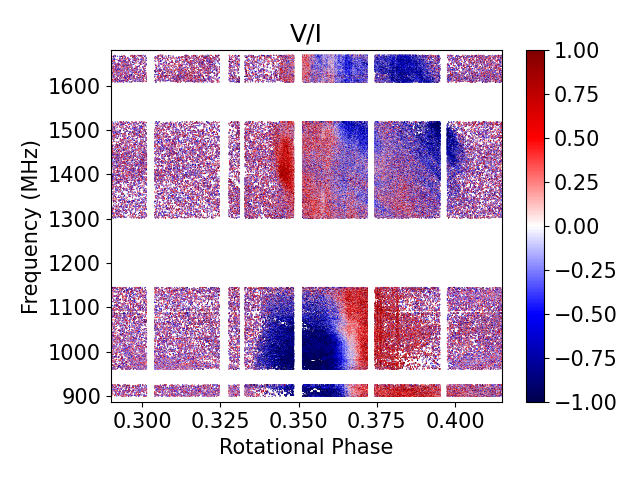}
    \includegraphics[width=0.325\textwidth]{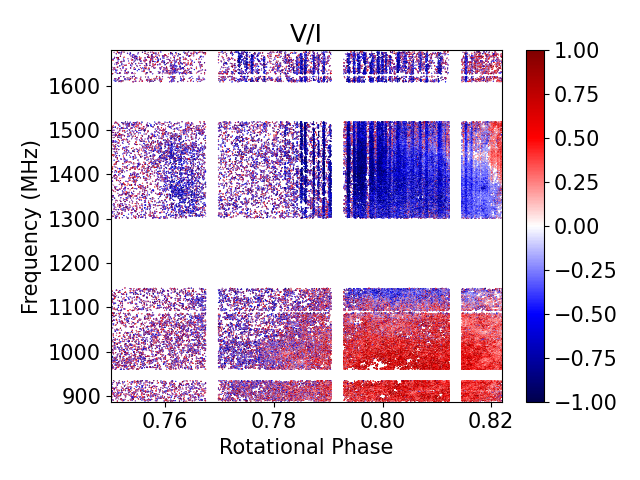}
    \includegraphics[width=0.325\textwidth]{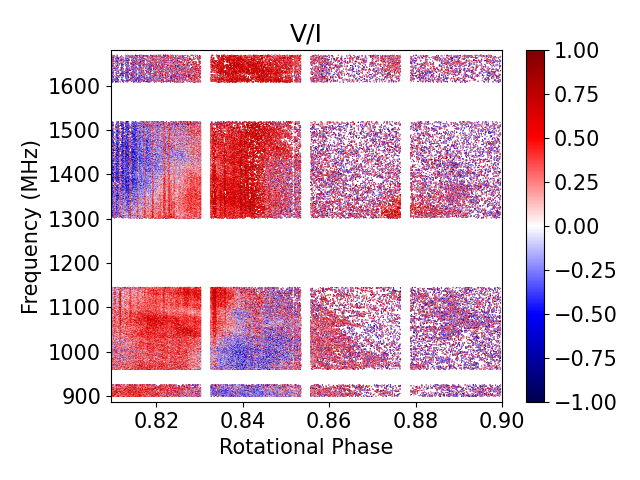}    
    \caption{The dynamic spectra of the primary pulses (see Figure \ref{fig:Lband_LC}). From top to bottom respectively: $LCP$, $RCP$, Stokes $I$, Stokes $V$ and $V/I$.
    }
    \label{fig:DS}
\end{figure*}

\begin{figure*}
    \centering
    \includegraphics[width=0.49\textwidth]{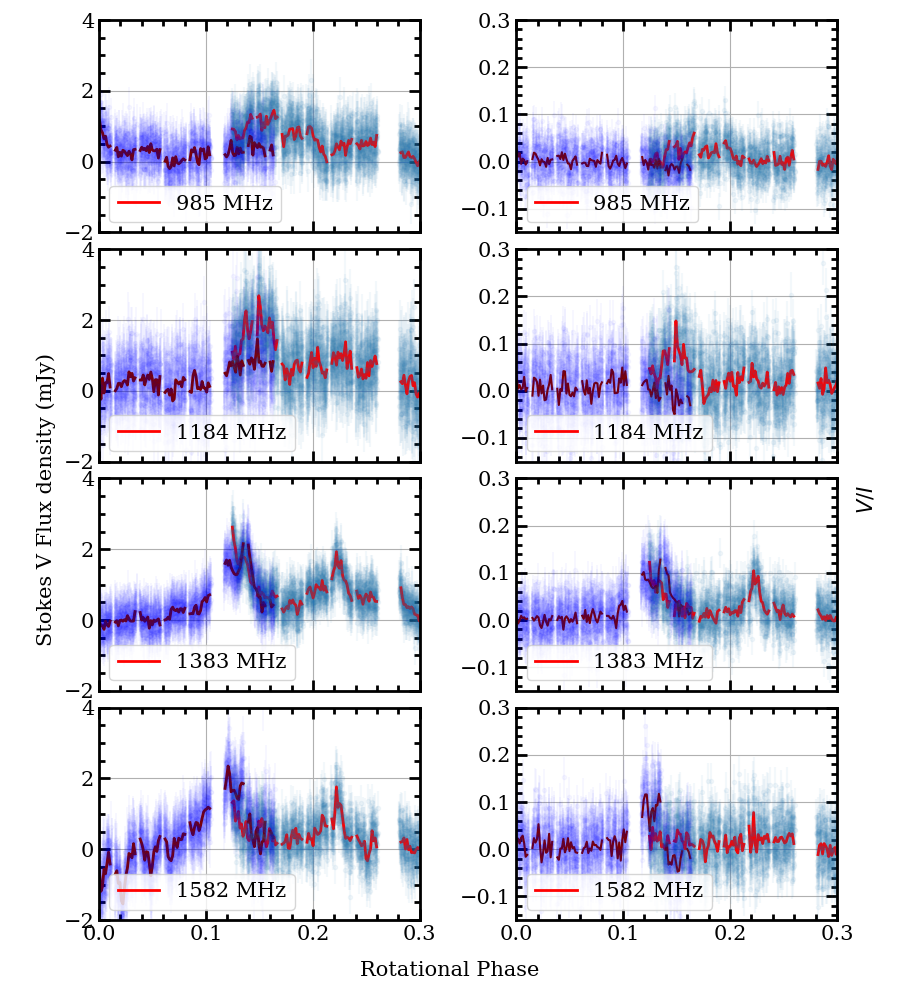}
    \includegraphics[width=0.49\textwidth]{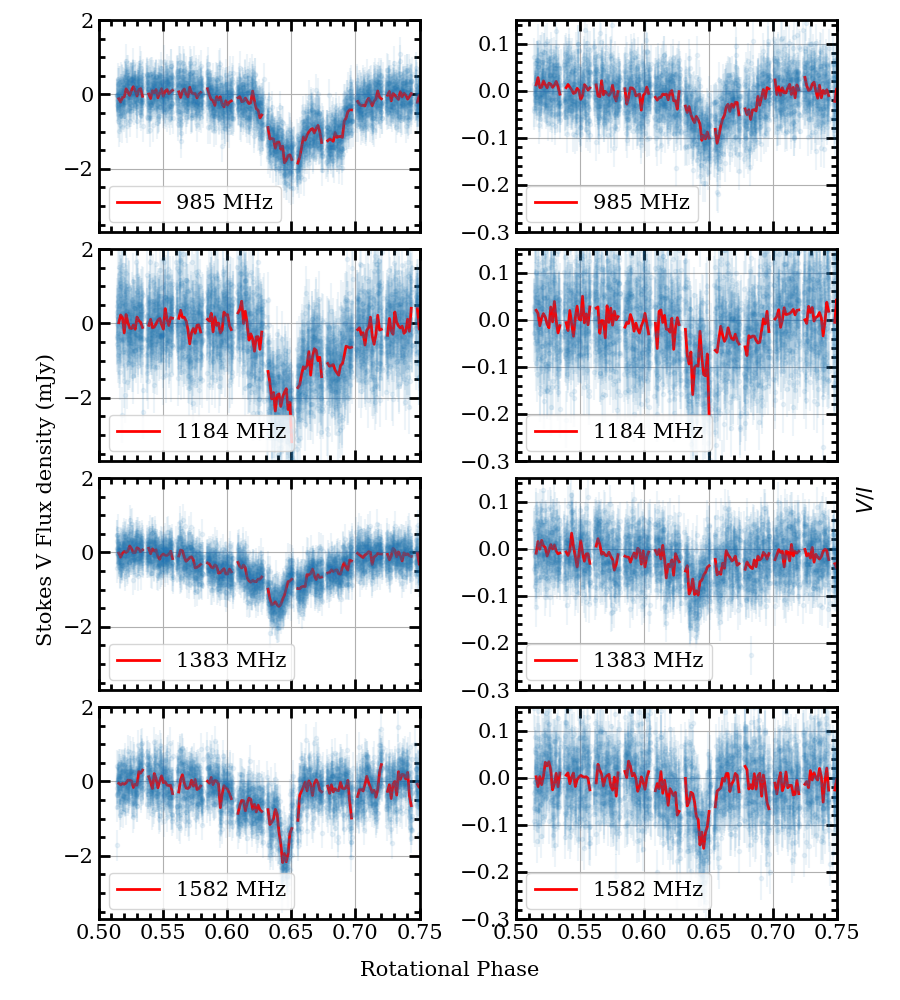}
    \caption{The frequency evolution of the lightcurves for the secondary enhancements. The markers correspond to data points at the native time resolution of 8 seconds, the solid lines are obtained by averaging over 15 data points (equivalent to a time resolution of 2 minutes). \textbf{Left:} The Stokes V lightcurves and the corresponding circular polarization of the secondary enhancement observed on rotational phase 0.1. The different shades of colors represent data acquired on different days of observation. \textbf{Right:} Same as the figure on the left, but for the enhancement observed around phase 0.65. The complete enhancement was observed on a single observing session.}
    \label{fig:secondary_stokesV_lc}
\end{figure*}

The frequency averaged $RCP$ (red) and $LCP$ (blue) lightcurves of HD\,142990 are shown in Figure \ref{fig:Lband_LC}. There are two secondary pulses, one in each circular polarization, in addition to the primary pulses. This makes HD\,142990 the third MRP to exhibit secondary pulses \citep[after CU\,Vir and HD\,12447,][]{das2021,das2022}.
We also searched for linearly polarized emission, but did not detect any. 
The $4\sigma$ upper limits for both $Q/I$ and $U/I$, when averaged over the full band and integrated for one hour covering the peak of the primary pulses, are $\approx 3\%$.

One important difference between the lightcurve obtained from the 2023 data (this paper) with that obtained from the 2019 data \citep{das2019a,das2023} is the arrival phases of the primary pulses. In 2019, the primary pulses were observed around the rotational phases 0.28 and 0.75, but in 2023, we observed the pulses around the phases 0.36 and 0.82. 
However, the observed differences are within the uncertainties (0.13 phase) associated with the rotational phases calculated for the 2023 data (\S\ref{sec:data}).

In addition to obtaining the lightcurves, we also extracted the dynamic spectra for $RCP$ and $LCP$. Only the primary pulses are detected in the dynamic spectra. The dynamic spectra for the two sets of primary pulses are shown in Figure \ref{fig:DS}. 
The pulse characteristics observed in lightcurves (Figure \ref{fig:Lband_LC}) and the dynamic spectra (Figure \ref{fig:DS}) are described in the following subsections.

\subsection{Secondary enhancements}\label{subsec:secondary_pulses}
The primary motivation for observing the star over full rotation cycle was to search for additional enhancements in the lightcurves (other than those observed close to the magnetic nulls). Our observations confirm the presence of secondary enhancements in the lightcurves at least over the frequency range of MeerKAT L-band (935--1630 MHz). 
These enhancements are observed in-between the primary pulses, the LCP is observed at phases $\approx 0.6-0.7$ and the RCP is observed at phases $\approx 0.10-0.25$.
The separation between the two secondary enhancements ($\approx 0.5$ phase) is similar to that between the two sets of primary pulses.
The LCP enhancement was entirely observed on Day 1, but the one in RCP was partly observed on Day 2 (phases 0.12--0.30) and partly on Day 3 (phases 0.10--0.16). The fact that over the phase range common to Day 2 and Day 3, the secondary enhancement was present on both days, confirms that at least the RCP secondary enhancement is a persistent characteristic of the star's radio emission (a pulse).

In order to investigate the polarization properties of the secondary enhancements and their evolution with frequencies, we construct the Stokes $V$ and $V/I$ lightcurves from the $RR$ and $LL$ visibilities by averaging over two spectral windows ($\approx 200$ MHz). The results are shown in Figure \ref{fig:secondary_stokesV_lc}. 
We find that the enhancements are broadband in nature. This aspect, combined with their much weaker flux densities, explain their non-detections in the dynamic spectra.
For both secondary enhancements, the observed circular polarization ($|V/I|$) is $\sim 10-20\%$. The enhancement observed around phase 0.1 (left in Figure \ref{fig:secondary_stokesV_lc}) exhibits positive circular polarization and the one around phase 0.65 (right in Figure \ref{fig:secondary_stokesV_lc}) exhibits negative circular polarization throughout the observing band.

In case of the LCP secondary enhancement (right in Figure \ref{fig:secondary_stokesV_lc}), the profile clearly evolves from a broader, double-peaked structure at lower frequencies (e.g., the top-most panel at 985 MHz) to a narrower, single-peaked structure at higher frequencies (e.g. the bottom-most panel at 1582 MHz). In contrast, the RCP secondary enhancement (left in Figure \ref{fig:secondary_stokesV_lc}) exhibits a clear double-peaked profile at higher frequencies, which is much less pronounced at lower frequencies. Interestingly, at lower frequencies, this pulse was stronger on Day 2 than that on Day 3 over the common rotational phase range (the discrepancy in flux densities remain even after consider a 10\% flux density offset due to observation on different days). On Day 3, the peak flux density of the pulse increases with increasing frequencies until 1383 MHz (maroon curves in the left of Figure \ref{fig:secondary_stokesV_lc}), but on Day 2, such a rise in flux density was observed only between 985--1184 MHz (red curves in the left of Figure \ref{fig:secondary_stokesV_lc}).

Based on Figure \ref{fig:secondary_stokesV_lc}, we infer that the secondary enhancements extend beyond the frequency range covered by the L-band observations. There is a hint of an increase in circular polarization fraction with increasing frequency for the pulse around phase 0.1. Although the circular polarization is much lower \citep[comparable to that observed for incoherent gyrosynchrotron emission above $\sim 10$ GHz, e.g.][]{leto2017}, ECME remains the favoured mechanism due to the following reasons:
\begin{enumerate}
    \item The enhancements are confined within a rotational phase window of width $<0.2$, which is significantly smaller than the timescale of rotational modulation observed in case of incoherent radio emission \citep[e.g.][]{lim1996,leto2018,das2022}.
    \item For the secondary enhancement that was partially covered on two different days, consistent handedness in circular polarization was observed (rotational phase range $\approx 0.1-0.15$ in Figure \ref{fig:secondary_stokesV_lc}). This is expected for ECME, but not for plasma emission, another coherent mechanism \citep{villadsen2019,callingham2021}.
    \item Such structures are expected to arise due to ECME from a star like HD\,142990 that has a very high obliquity.
\end{enumerate}
The last point will be discussed in more detail in \S\ref{subsec:simulation}.

\subsection{Primary enhancements}\label{subsec:primary_enhancement}
Although the primary pulses of the star have already been reported in past studies \citep{das2019a, das2023}, they did not cover the frequency range $800-1000$ MHz. In addition, the properties of these pulses had been studied only at coarse spectral and time resolutions \citep[$\geq 62.5$ MHz and $\geq 1$ minute respectively,][]{das2023}. This paper reports the pulse structures at much higher time and spectral resolutions ($8$ seconds and $418$ kHz, respectively) for the first time.

In Figure \ref{fig:DS}, we show the dynamic spectra for the primary pulses observed on the three days. The pulse close to null 1 (phase $\approx 0.25$) was completely observed within a single observation session (left column in Figure \ref{fig:DS}, Day 2), whereas the pulse close to null 2 (phase $\approx 0.66$) was partially covered on two different days (middle and right columns in Figure \ref{fig:DS}, Day 1 and 3). From top to bottom, the dynamic spectra correspond to emission in $LCP$, $RCP$, Stokes $I$, Stokes $V$ and $V/I$ respectively. These high time and spectral resolution plots revealed several unique characteristics of the phenomenon that were missed in past studies. These are described in the following subsections.

\subsubsection{First observation of fine structures for ECME from magnetic hot stars}\label{subsubsec:fine_structures}
\begin{figure*}
    \includegraphics[width=0.99\textwidth]{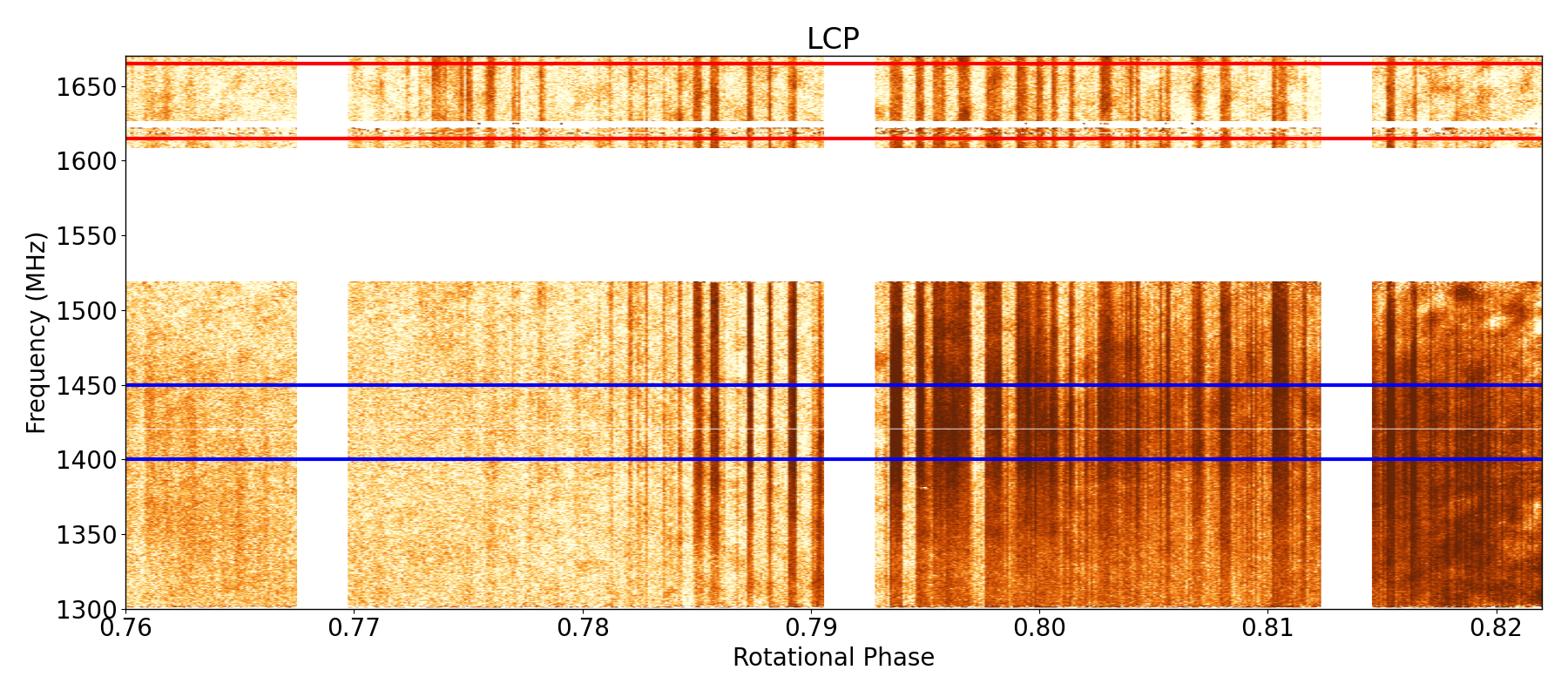}
    \includegraphics[width=0.99\textwidth]{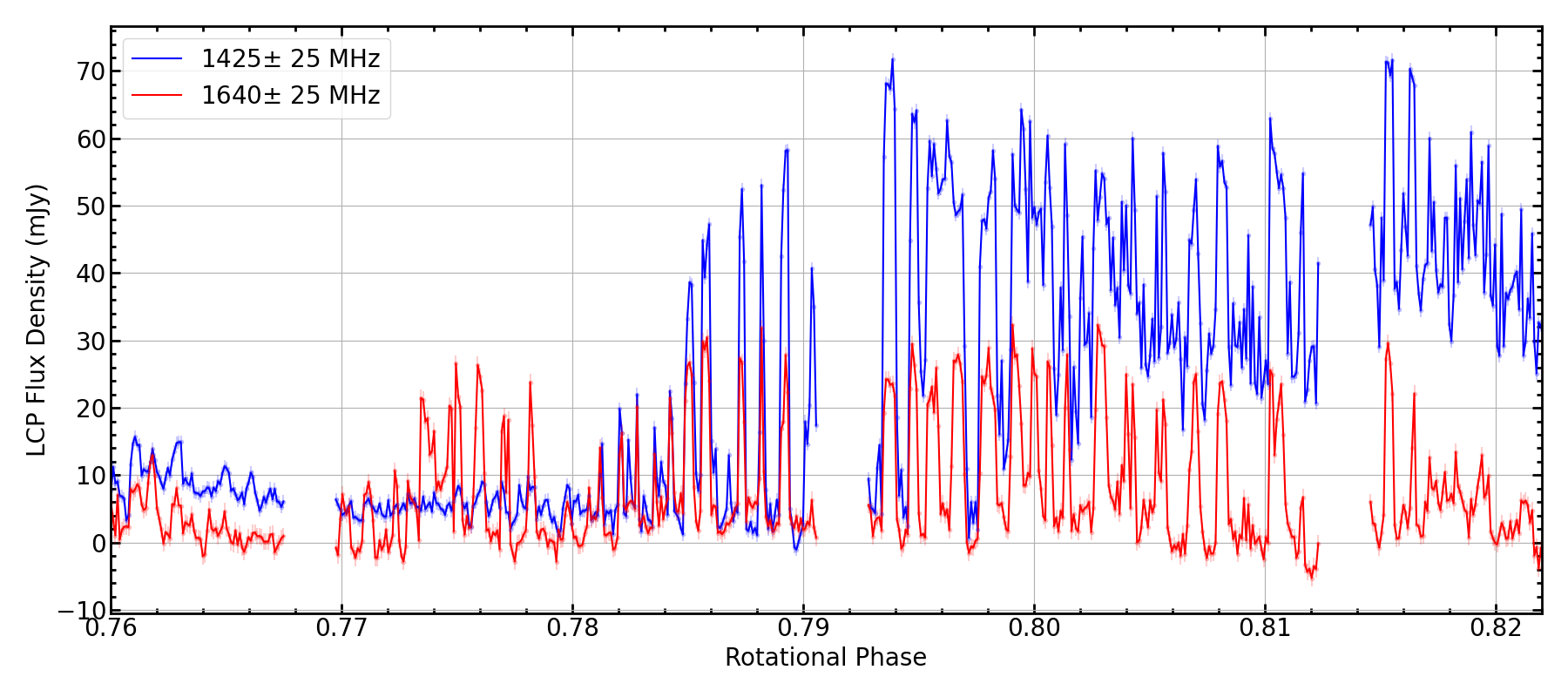}
    \caption{The fine structures seen in LCP pulses on Day 1 of our observation. \textbf{Top:} The dynamic spectrum zoomed over the frequency range of 1300--1670 MHz. \textbf{Bottom:} The lightcurves extracted from the dynamic spectrum (top panel) at two central frequencies of 1425 (blue) and 1640 (red) MHz by averaging over a bandwidth of 50 MHz in each case. The averaging frequency ranges are also indicated by the horizontal lines on the top panel.}
    \label{fig:LCP_fine_structures_day1}
\end{figure*}

\begin{figure*}
    \includegraphics[width=0.99\textwidth]{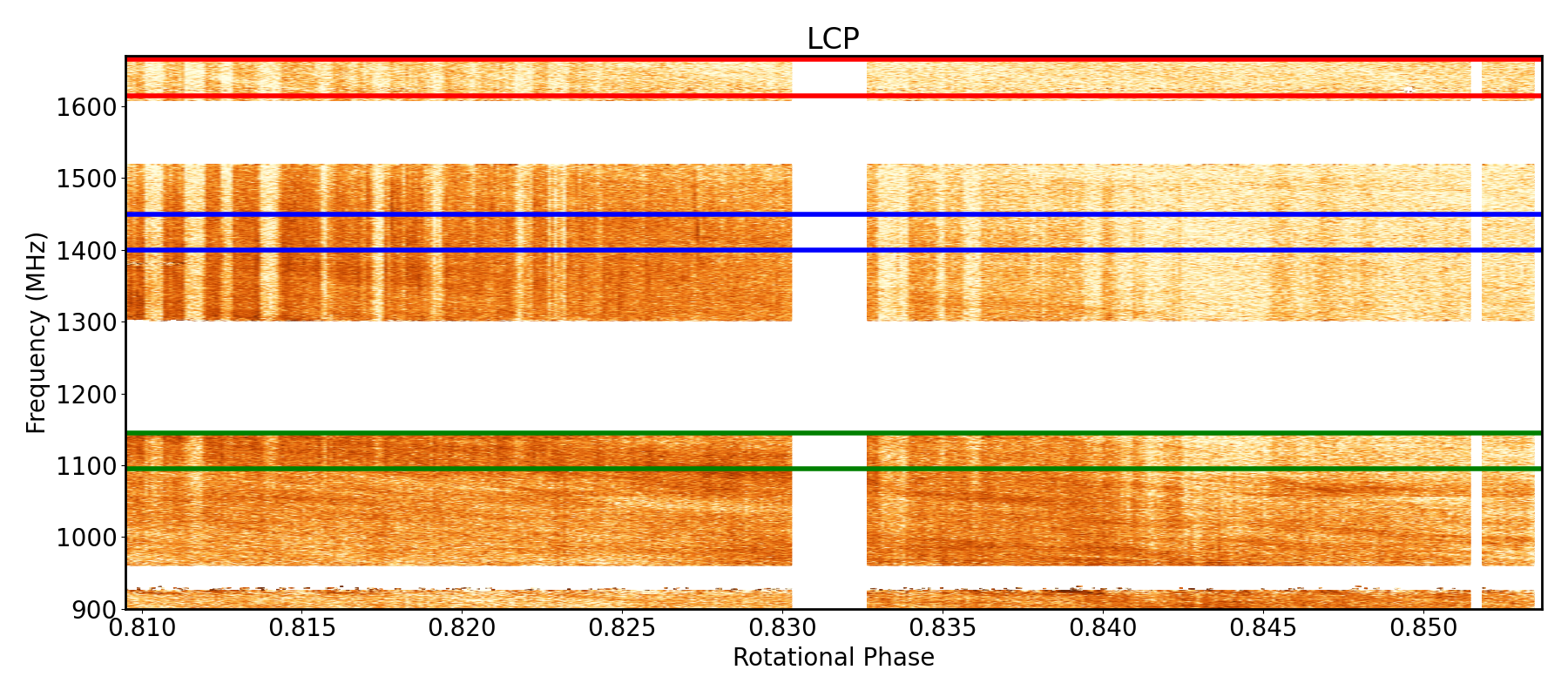}
    \includegraphics[width=0.99\textwidth]{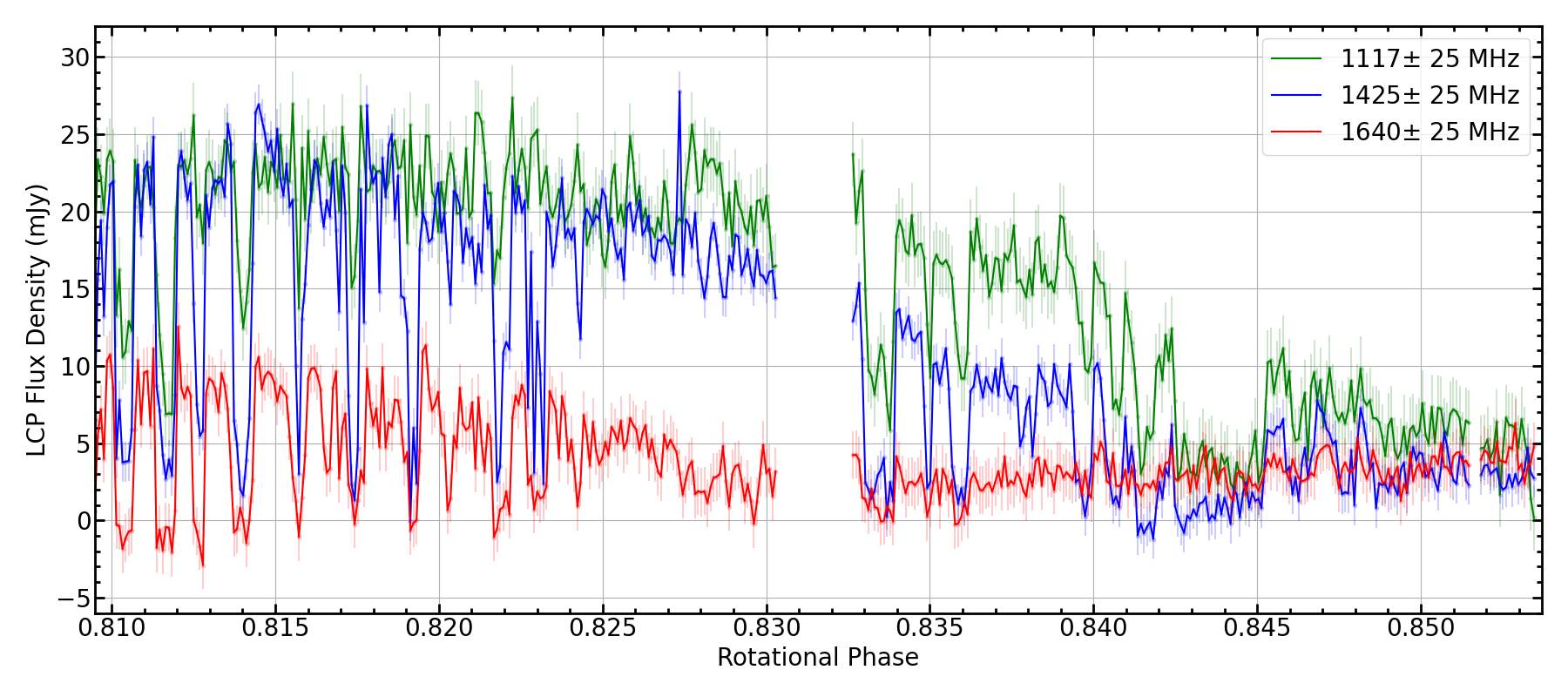}
    \caption{The fine structures seen in LCP pulses on Day 3 of our observation. \textbf{Top:} The dynamic spectrum zoomed over the frequency range of 1300--1670 MHz. \textbf{Bottom:} The lightcurves extracted from the dynamic spectrum (top panel) at three central frequencies of 1117 MHz (green), 1425 (blue) and 1640 (red) MHz by averaging over a bandwidth of 50 MHz in each case. The averaging frequency ranges are also indicated by the horizontal lines on the top panel.}
    \label{fig:LCP_fine_structures_day3}
\end{figure*}

\begin{figure*}
\centering
    \includegraphics[trim={0 0 4cm 0},clip, width=0.8\textwidth]{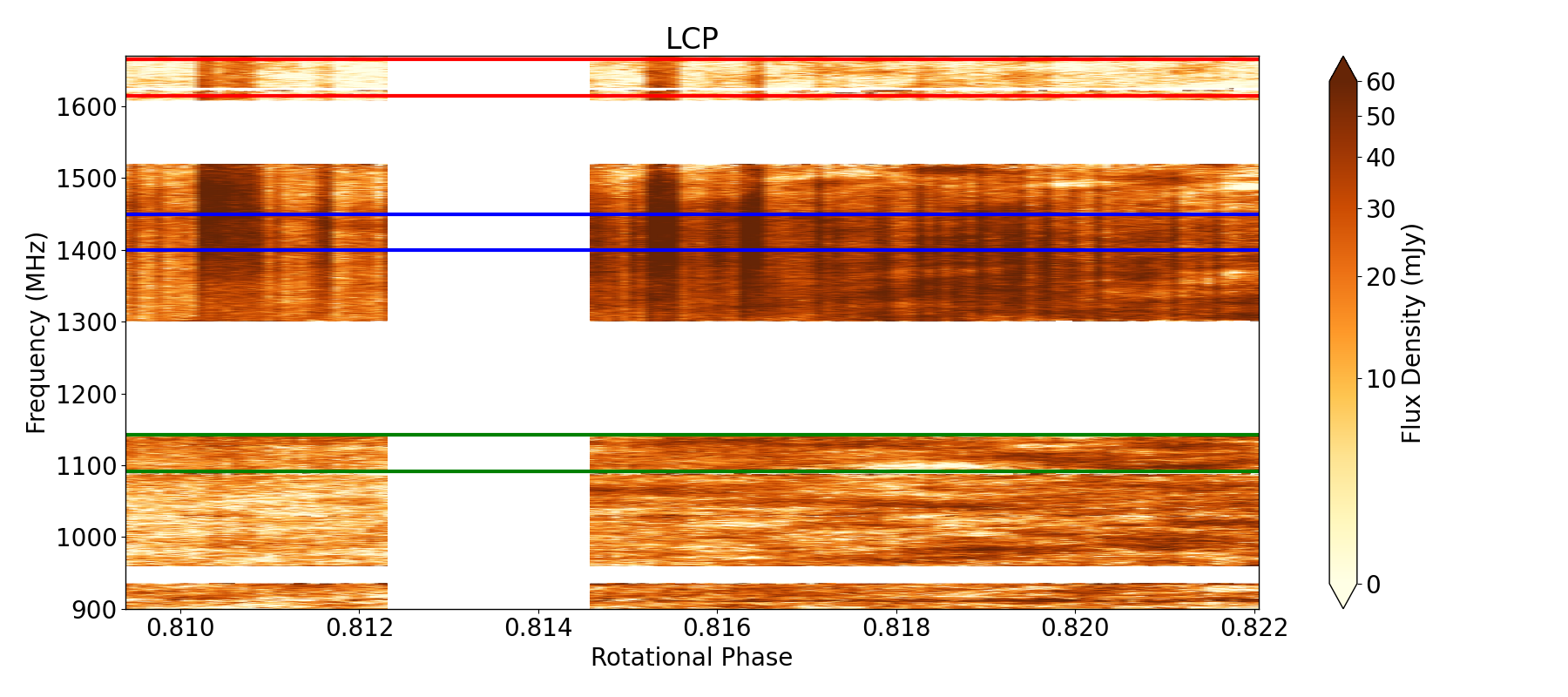}
    \includegraphics[trim={0 0 4cm 0},clip, width=0.8\textwidth]{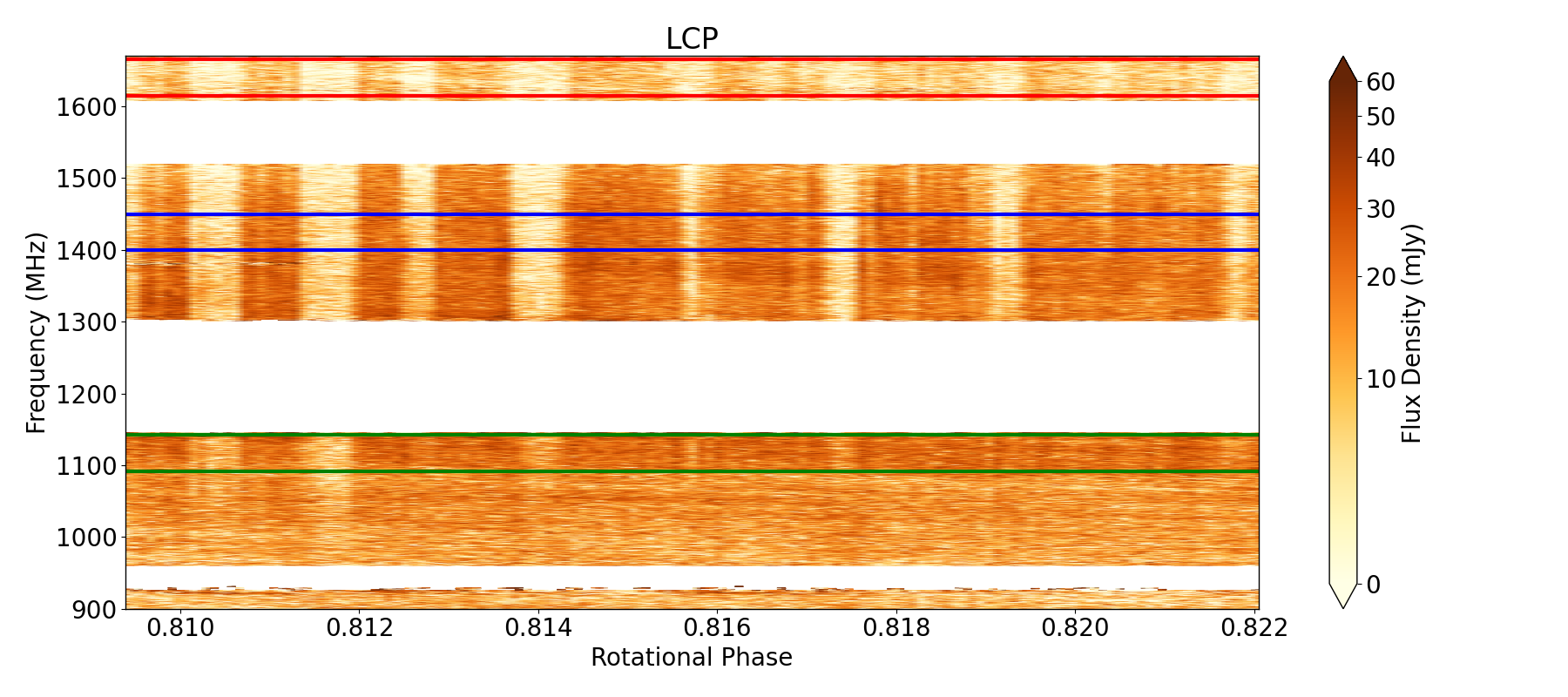}
    \includegraphics[width=0.8\textwidth]{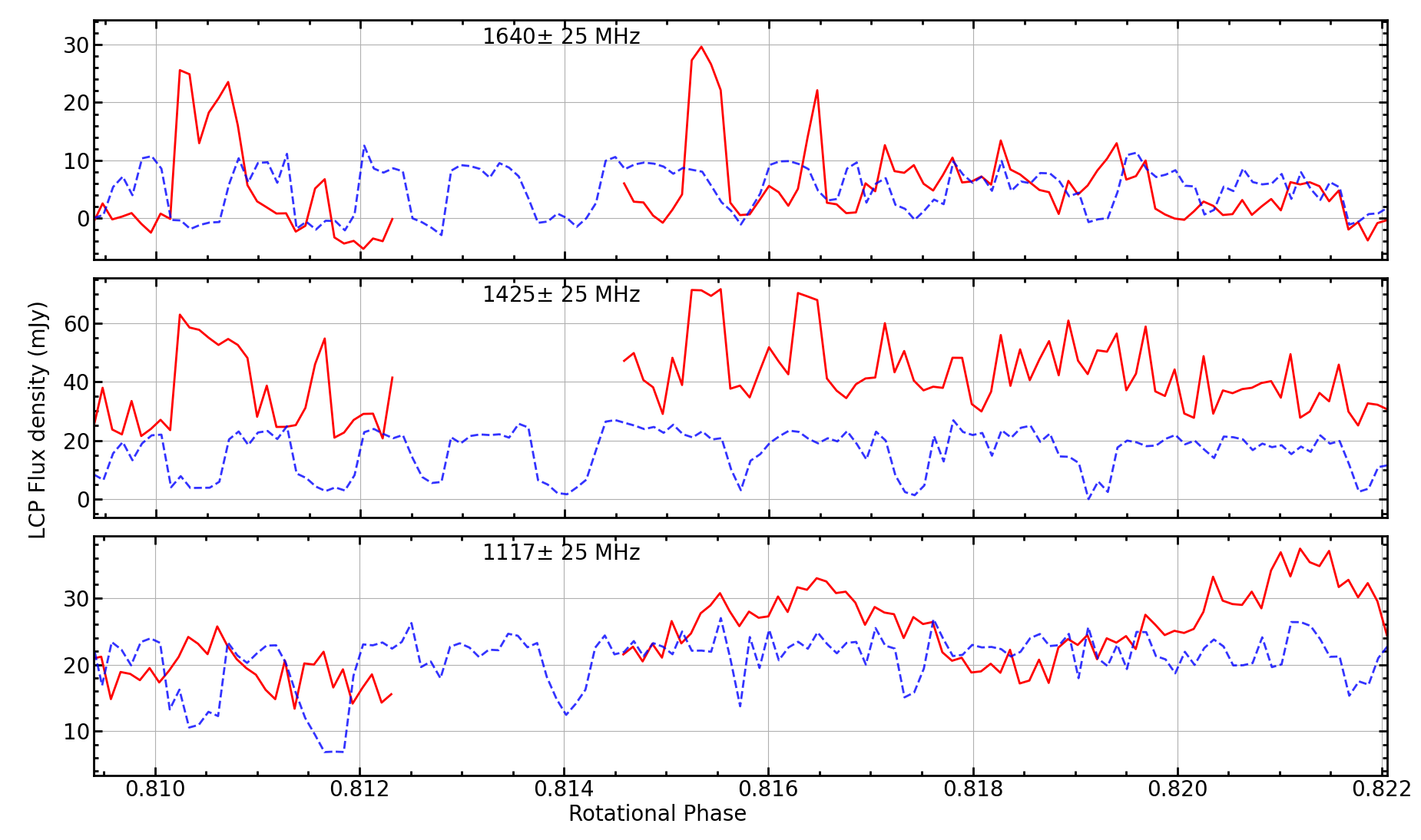}
    \caption{Comparison of the LCP fine structures over the rotational phases common to observations on Day 1 and Day 3. The top two panels show the dynamic spectra from Day 1 and Day 3 respectively. The bottom panel compares the corresponding lightcurves at three frequencies (marked in the dynamic spectra) with red solid curves for Day 1 and blue dashed curves for Day 3.
    }
    \label{fig:LCP_fine_structures_common_phi_range}
\end{figure*}

The most notable feature that we discover in the dynamic spectra are the fine structures observed in LCP over the rotational phase range $\approx 0.76-0.86$ (first row, middle and right columns in Figure \ref{fig:DS}). The LCP pulse over this phase range consists of short spikes of duration as small as 8 seconds (the time resolution of the data), suggesting that a higher time resolution observation might reveal finer structures in the emission. These structures were seen on both days of our observation covering the LCP pulse, showing that these fine structures are stable characteristics of the LCP pulse.

In Figure \ref{fig:LCP_fine_structures_day1}, the fine structures observed on Day 1 are shown by zooming over the frequency range of 1300--1670 MHz (top). In order to study the spectral properties of these structures, we extract the lightcurves by averaging over 50 MHz around two frequencies: 1425 MHz and 1640 MHz (bottom panel). Comparing these lightcurves, we find that at lower frequencies, the spikes are superimposed on a smoother `envelope' of emission following the rotational phase of $\approx 0.80$ (bottom panel). This envelope vanishes completely at 1640 MHz (can be seen from the flat baseline, bottom panel of Figure \ref{fig:LCP_fine_structures_day1}). Thus, at 1640 MHz, the LCP enhancement is entirely composed of these spikes. For phases $<0.80$, the envelope is not prominent even for the lower frequency. 
While most of the spikes are present in both frequencies, their spectra evolve significantly as a function of rotational phase. At phases $\lesssim 0.77$, the spikes are barely recognizable and the emission at 1425 MHz is dominated by the underlying smooth envelope. Between 0.77--0.78 phases, the spikes are prominent, but only at the higher frequency. Between 0.780--0.785, spikes are prominent in both frequencies and are of similar strengths. Between 0.785--0.80 phase, the spikes are clearly much brighter at the lower frequency. Beyond phase 0.80, it is not possible to disentangle the spikes from the background emission at the lower frequency and hence we cannot compare the spikes at the two frequencies. 
Note that, within our spectral and temporal resolutions, we do not detect drift of the individual spikes.

The fine structures observed on Day 3 are shown in Figure \ref{fig:LCP_fine_structures_day3}. In this case, we show the lightcurves at three representative frequencies: 1117 MHz, 1425 MHz and 1640 MHz. The peak flux densities are much lower on this day, as can be seen from the lightcurves over the phases 0.81--0.82 (also covered on Day 1, Figure \ref{fig:LCP_fine_structures_day1}). Here, the spikes are the weakest at the highest frequency and non-recognizable beyond phase 0.83. Until phase 0.83, the spikes at 1425 MHz and 1117 MHz have similar peak flux densities, however, this could also be due to a stronger background emission at lower frequencies. Above phase 0.83, the emission is the strongest at the lowest frequency till around phase 0.842. The region between phases $\approx 0.843-0.845$ is devoid of spikes (and also of background emission, see the top panel of Figure \ref{fig:LCP_fine_structures_day3}). Beyond this phase, weak spikes emerge again at the two lower frequencies. The spikes become non-detectable beyond phase 0.85.

As mentioned already, the rotational phases covered by Day 1 and Day 3 overlap over the phases 0.809--0.822. The LCP fine structures from the two days over this phase range are compared in Figure \ref{fig:LCP_fine_structures_common_phi_range}. The main points that we note from this figure are:
\begin{enumerate}
    \item The fine structures observed on the two days are not identical. This suggests that the spikes vary in intensity \citep[known to occur for ECME, e.g.][]{trigilio2011,das2021}.
    \item Both the spikes and the underlying smooth envelope are much weaker on Day 3 (e.g. compare the lightcurves at 1640 and 1425 MHz in Figure \ref{fig:LCP_fine_structures_common_phi_range}). This could happen when they have a common driver or their driving mechanisms are related.
    \item Up to the rotational phase of $\approx 0.812$, the enhancements observed on the two days appear to be anti-correlated (crests on the lightcurves of one day coincide with troughs on the lightcurves of other day; e.g. see the middle panel of Figure \ref{fig:LCP_fine_structures_common_phi_range}). The underlying reason(s) is not known at the moment.
    \item On Day 3, the minimum observable frequency for the spikes is $\lesssim 1117$ MHz, but on Day 1, the spikes exhibit a cut-off at a frequency $\gtrsim 1150$ MHz (e.g., see the frequency ranges $1117\pm 25$ MHz, enclosed by the green horizontal lines in Figure \ref{fig:LCP_fine_structures_common_phi_range}). This could be a result of the fact that on Day 1, the background envelope is much brighter than that on Day 3, so that the weaker spikes are not distinguishable.
\end{enumerate}
The last point can be extrapolated to explain why we see these prominent fine structures only for the LCP pulse near null 2. Among the four pulses (two LCP and two RCP), the LCP pulse near null 2 exhibits the lowest upper cut-off frequency. Its peak flux density spectrum declines steeply beyond $\approx 1500$ MHz reaching a cut-off at $\approx 2$ GHz \citep[see the bottom left panel of Figure 4 in][]{das2023}. On the other hand, the RCP pulse around the same magnetic null, for which the emission appears very smooth on the frequency-time plane (second row, third column in Figure \ref{fig:DS}), exhibits the highest upper cut-off frequency \citep[$\approx 3.3$ GHz,][]{das2023}. Thus, it is possible that the fine structures become detectable only close to the cut-off frequencies.

Faint spikes are also seen for $RCP$ pulse near null 1 between 0.36--0.39 over $\approx 900-1150$ MHz (Figure \ref{fig:RCP_DS_LC_day2}). These fine structures appear close to the edges of the primary pulse for which the underlying smooth emission is weaker. No such spikes are observed for the other primary $RCP$ pulse (null 2) or the $LCP$ pulse near null 1.

\subsubsection{Reversal in the drift direction of pulses}\label{subsubsec:drift_reversal}
\begin{figure*}
    \centering
    \includegraphics[width=0.99\textwidth]{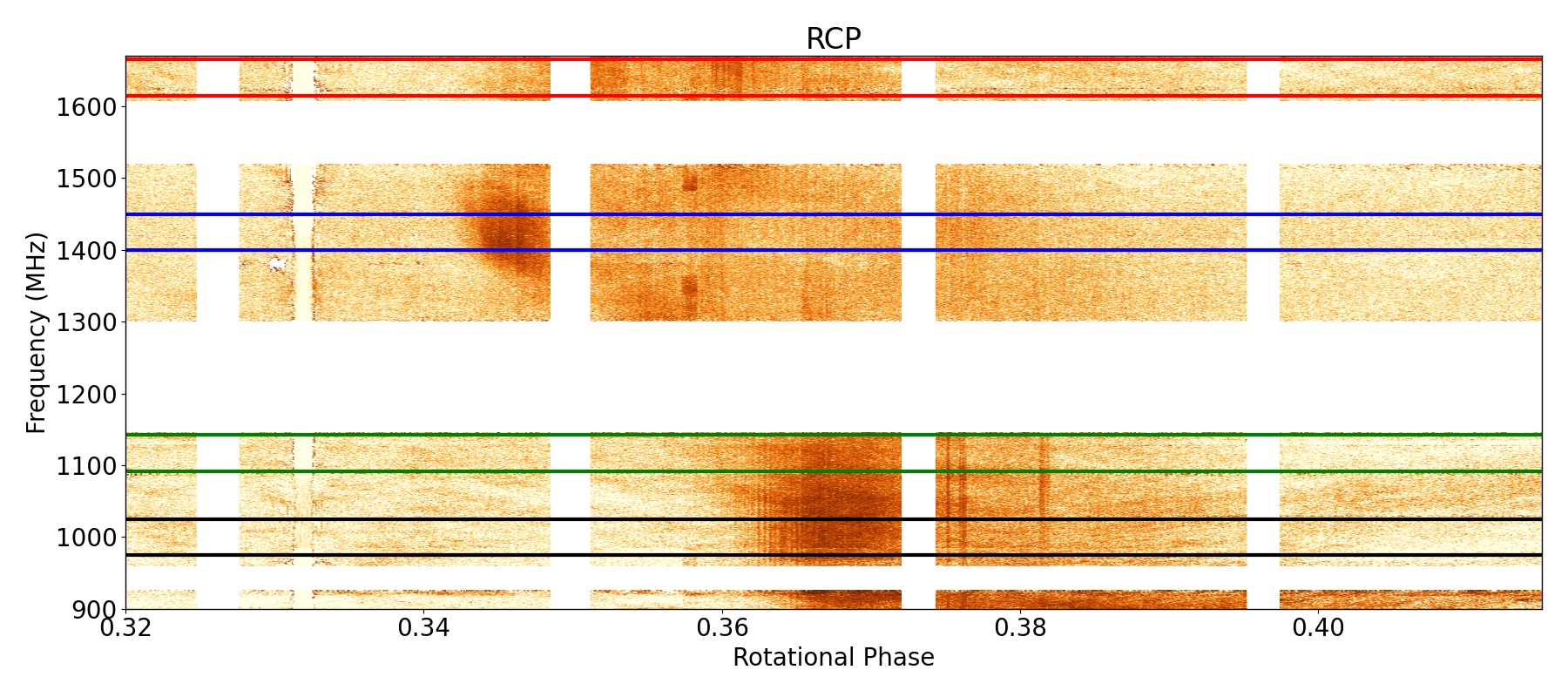}
    \includegraphics[width=0.99\textwidth]{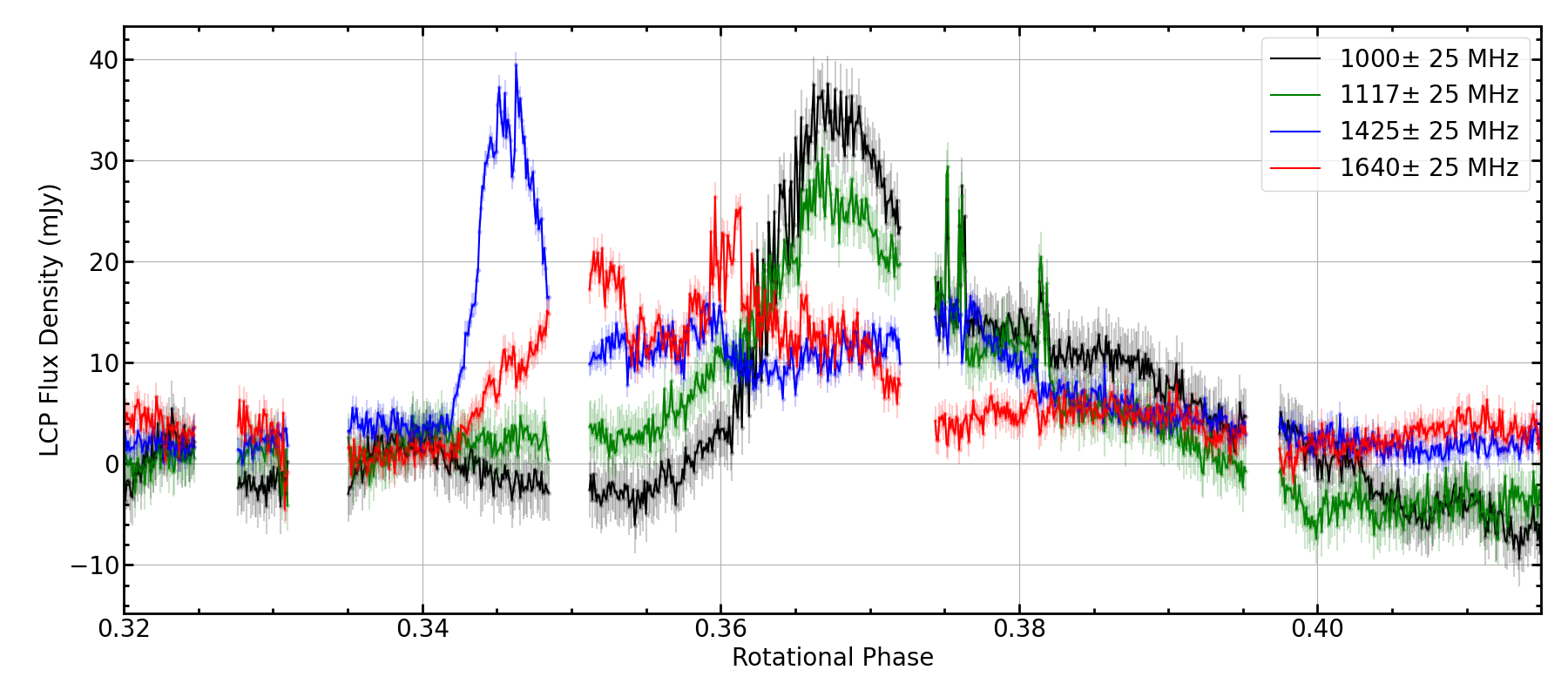}
    \caption{The primary $RCP$ pulses near null 1 observed on Day 2. The \textbf{top} panel shows the dynamic spectrum, with horizontal lines marking the frequency ranges chosen to extract lightcurves. These lightcurves are shown on the \textbf{bottom} panel}
    \label{fig:RCP_DS_LC_day2}
\end{figure*}
\begin{figure*}
    \centering
    \includegraphics[width=0.99\textwidth]{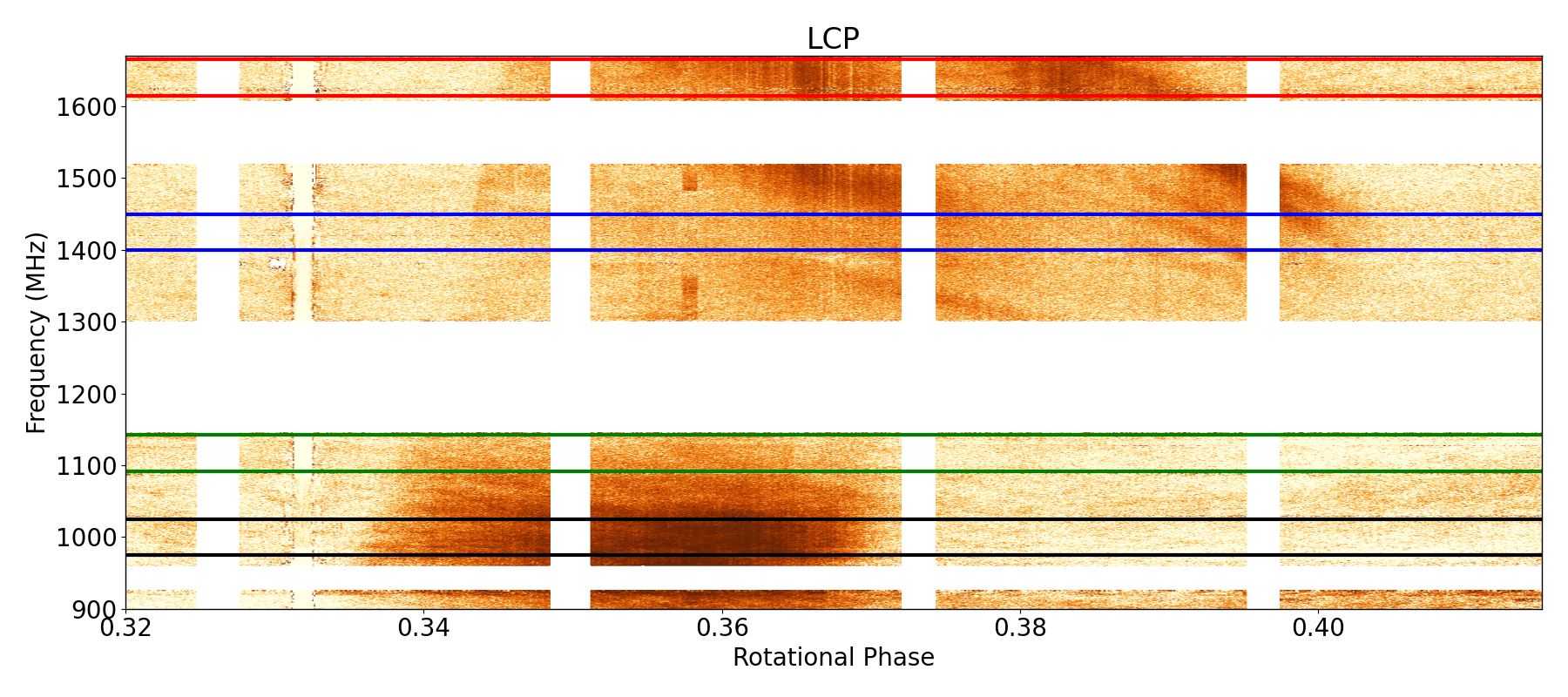}
    \includegraphics[width=0.99\textwidth]{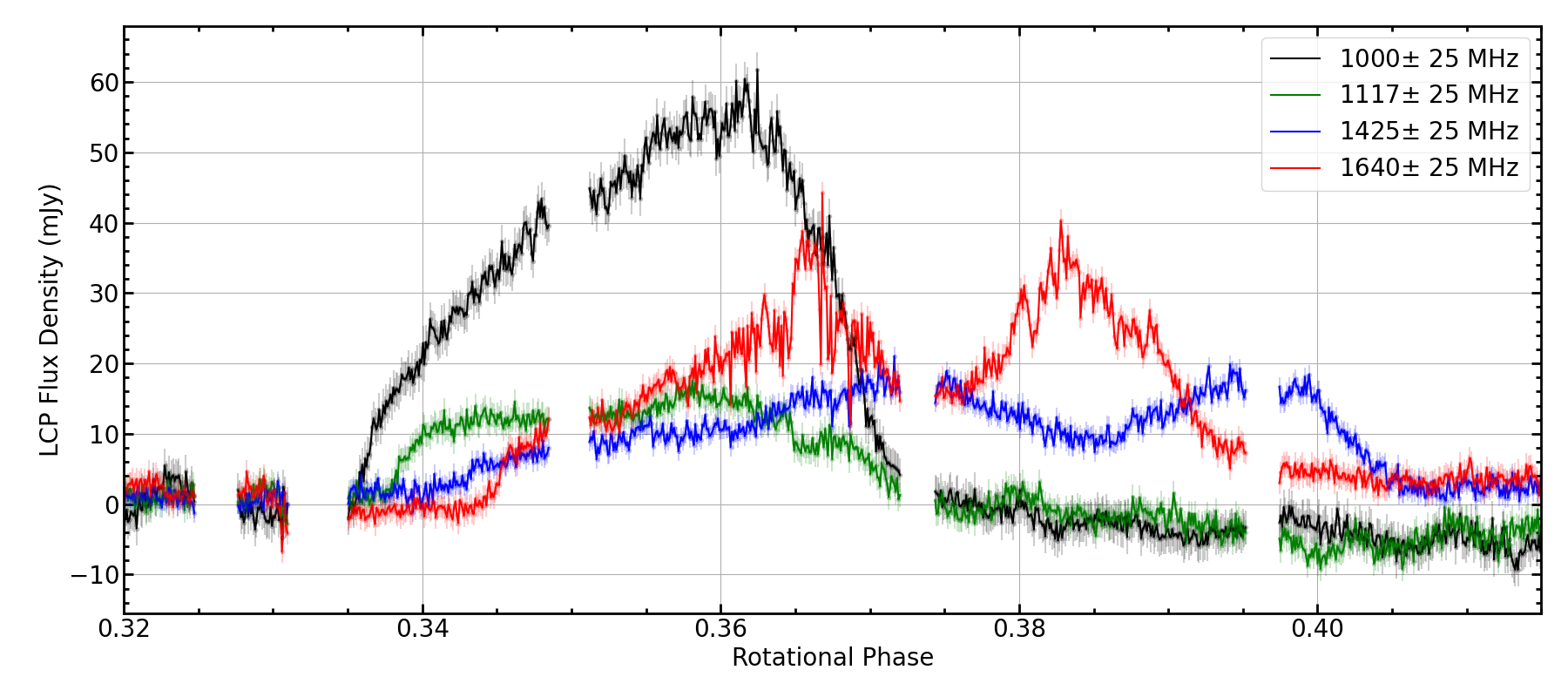}
    \caption{Same as Figure \ref{fig:RCP_DS_LC_day2} but for the primary $LCP$ pulse.}
    \label{fig:LCP_DS_LC_day2}
\end{figure*}
\citet{das2023} reported that the sequence of arrival of the $RCP$ and $LCP$ pulses reverses with increasing frequencies for the pulses around null 1 and speculated that a similar phenomenon occurs for the pulses around null 2. Our Stokes $V$ dynamic spectra confirm this phenomenon for the pulses observed around both magnetic nulls (see the fourth row in Figure \ref{fig:DS}). 

The dynamic spectra reveal a subtle difference between the drifting properties of the pulses around the two nulls (Figure \ref{fig:DS}). Around null 2, the reversal is caused by the fact that the $RCP$ ($LCP$) pulse continues to drift towards later (earlier) times with increasing frequencies throughout our observing band (second and third columns in Figure \ref{fig:DS}). In the ideal scenario, once the $RCP$ and $LCP$ pulse overlap, no further drifting should  occur with increasing frequencies \citep{trigilio2011,leto2016,das2020a}. This is clearly violated by the pulses observed around null 2.

The pulses around null 1 also violate the ideal scenario, but in a different manner. The $RCP$ pulse continues to drift towards earlier times with increasing frequencies up to around $\approx 1450$ MHz. Beyond that frequency, no further drifting is noticeable (Figure \ref{fig:RCP_DS_LC_day2}). This is consistent with the observation of \citet{das2023} that the $RCP$ pulse near null 1 appears over the same rotational phase range above $\approx 2$ GHz. For the corresponding $LCP$ pulse, \citet{das2023} reported that the drift direction reverses between 1.8--2.3 GHz. They also reported that the $LCP$ pulse separates into two sub-pulses at 1.8 GHz and one of the sub-pulses appears over the same rotational phase range for frequencies $\geq 1.8$ GHz, whereas the other component drifts in a direction opposite to that of $LCP$ pulse at lower frequencies between 1.8--2.3 GHz. From the MeerKAT data, we find that the $LCP$ pulse separates into two sub-pulses above 1400 MHz, both of which drift in a direction opposite to that of the lower frequency (single-peaked) $LCP$ pulse (Figure \ref{fig:LCP_DS_LC_day2}). Interestingly, between 1300--1400 MHz, we do not see any corresponding primary enhancement, but there are hints of multiple narrower and much fainter enhancements drifting in the same direction as the higher frequency $LCP$ sub-pulses. Thus the $LCP$ pulses observed above 1400 MHz may not be related to that observed below 1300 MHz.

\citet{das2023} also reported strong evolution in pulse-profiles with frequencies, especially for the pulses observed near null 2. For the pulses near null 1, significant evolution of pulse-profile was observed only above $\approx 1.6$ GHz \citep[see figures 8 and 9 of][]{das2023}. The MeerKAT data show that even below 1.6 GHz, the pulses around null 1 undergo significant evolution in their profiles and drift direction. In \S\ref{subsec:simulation}, we will examine whether or not these anomalous behaviors can be explained by propagation effects in the magnetospheres of a star with a highly oblique magnetic axis.

\subsubsection{Circular polarization of primary pulses}\label{subsubsec:stokesV_primary_pulses}
\begin{figure*}
    \centering
    \includegraphics[width=0.45\textwidth]{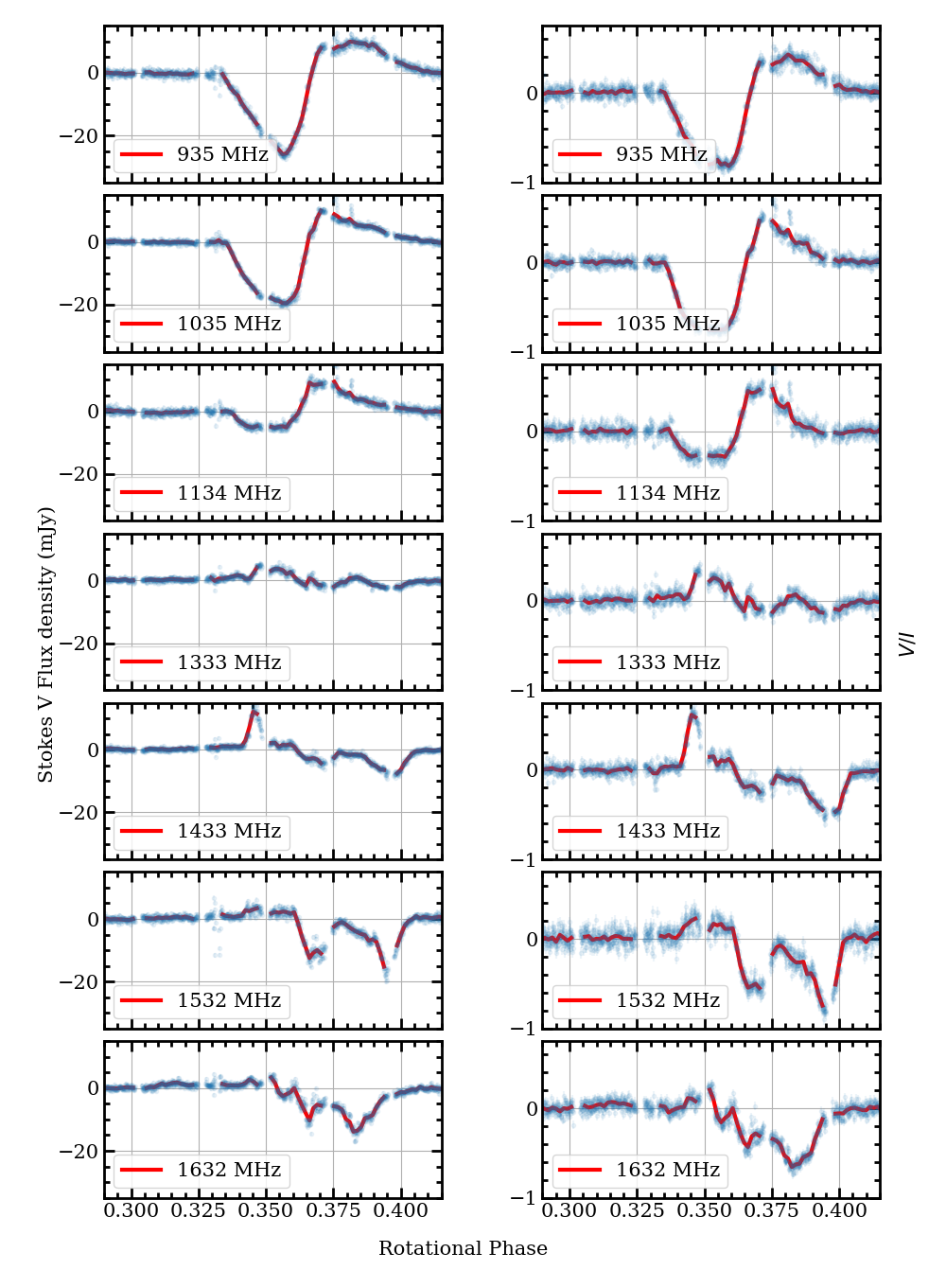}
    \includegraphics[width=0.45\textwidth]{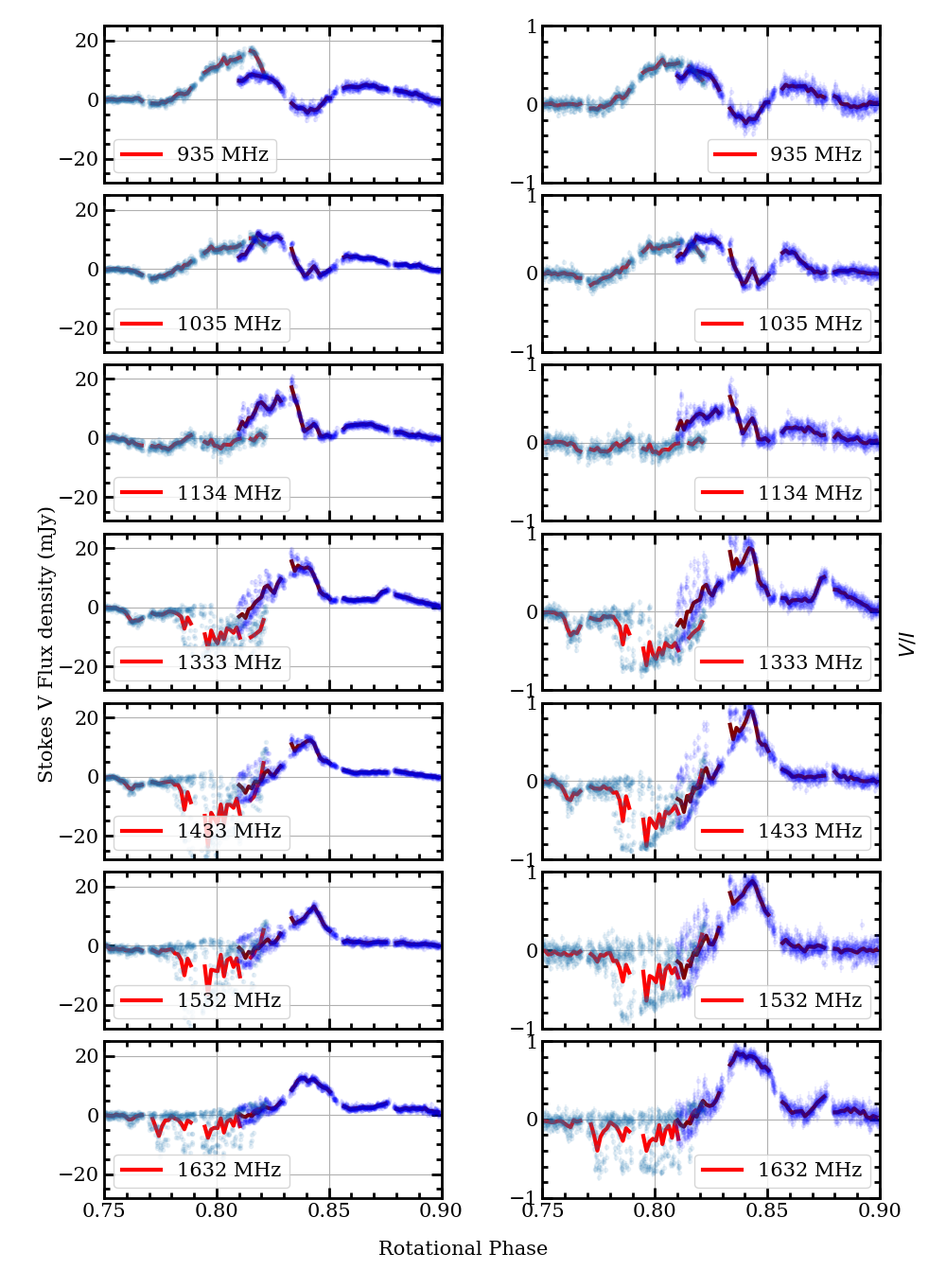}
    \caption{The Stokes $V$ and $V/I$ lightcurves for the primary pulses as a function of frequencies. The \textbf{left} plot shows the pulses observed around null 1 and the \textbf{right} plot shows the pulses observed close to null 2. The different shades of colors on the right plot indicate data from two different days. The markers correspond to data points at the native time resolution of 8 seconds, the solid lines are obtained by averaging over 15 data points (equivalent to a time resolution of 2 minutes).}
    \label{fig:primary_pulse_pol_properties}    
\end{figure*}

The Stokes $V$ and $V/I$ lightcurves (averaged over $\approx 100$ MHz) of the primary pulses are shown in Figure \ref{fig:primary_pulse_pol_properties}. For the pulses near null 1 (left plot in Figure \ref{fig:primary_pulse_pol_properties}), the percentage circular polarization varies between $\approx +60\%$ to $>90\%$. The Stokes $V$ profile transitions from the characteristic `$S$' shape at lower frequencies to a more complex shape at around a frequency of 1333 MHz (at which, the Stokes $V$ signal is the weakest, $\lesssim 20\%$). In case of the pulses close to null 2, the $RCP$ pulse is known to exhibit double peaks at frequencies $\lesssim 1.4$ GHz \citep{das2019a,das2023}. Although this aspect is not easily visible in the dynamic spectra, it can be clearly seen in the $V$ and $V/I$ profiles over 935--1333 MHz (right plot in Figure \ref{fig:primary_pulse_pol_properties}). The percentage circular polarization varies between $\pm \approx 90\%$. Between 1333--1632 MHz, $\approx 100\%\, RCP$ pulse is observed between phase 0.82--0.85.

As mentioned already, the complete coverage of the pulses near null 2 was obtained by observing on two different days: Day 1 and 3. The peak flux densities in $RCP$ and $LCP$ were not identical on the two days, which is also reflected in the Stokes $V$ lightcurves obtained from the two days over the common rotational phase range. The disagreement is the most prominent at the lowest frequency of 935 MHz and vanishes above 1.3 GHz. 



\section{Discussion}\label{sec:discussion}
The two main discoveries that this paper reports are the production of secondary enhancements/pulses and fine structures of coherent radio emission. In the next subsection, we discuss the origin of secondary enhancements due to propagation effects in a stellar magnetosphere with strong density gradients. This is followed by a discussion on possible explanations behind the observed fine structures.

\subsection{Secondary enhancements: consequences of magnetospheric propagation effects?}\label{subsec:simulation}
\begin{figure}
    \centering
    \includegraphics[width=0.45\textwidth]{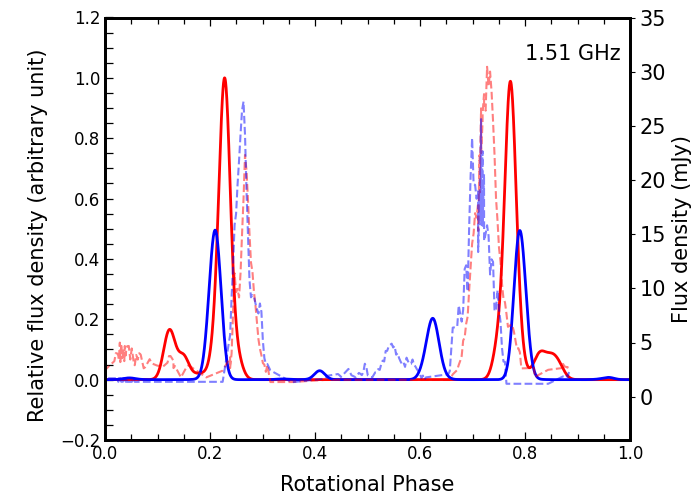}
    \includegraphics[width=0.45\textwidth]{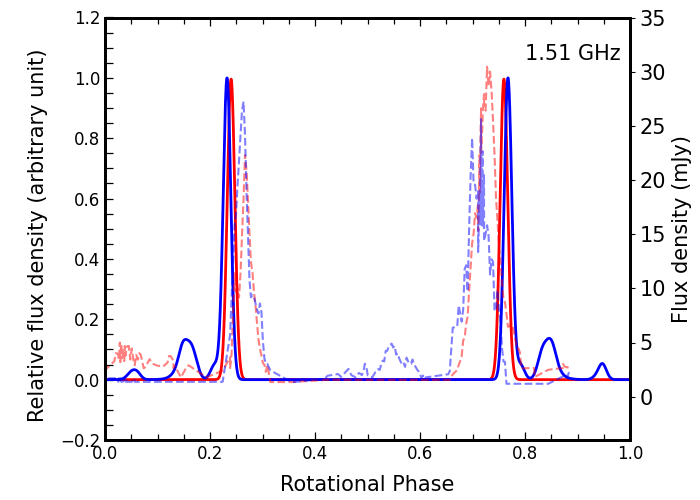}
    \caption{The simulated lightcurves at 1.5 GHz obtained using the 3D-framework of \citet{das2020a}, shown in solid curves, correspond to the Y-axes on the left hand side. The values of different parameters used in the simulation are given in \S\ref{subsec:simulation}. The red and blue curves represent RCP and LCP respectively. The observed RCP and LCP lightcurves are also overlaid in red and blue dashed curves respectively (correspond to the Y-axis on the right hand side). Note that the rotational phases of the observed lightcurves are shifted by $0.1$ phase to minimize the offsets between the observed and simulated primary pulses. 
    The simulated lightcurves on the top panel are obtained using an inclination angle of $i=36^\circ$ (provides closest resemblance to the observed lightcurves in terms of number and relative location of the secondary enhancements with respect to the primary pulses), and those on the \textbf{bottom} panels are obtained using an inclination angle of $55^\circ$ \citep[reported for HD\,142990,][]{shultz2019c}. Note that the zero of the rotational phases corresponds to the maximum negative \bz~similar to the convention adopted by \citet{shultz2019_0}.}
    \label{fig:best_case_sim_lc}
\end{figure}
In order to investigate whether secondary enhancements can arise due to refractions suffered in a stellar magnetosphere with highly complex plasma density distribution \citep[a result of large obliquity,][]{townsend2005}, 
we use the 3D framework of \citet{das2020a} to simulate ECME lightcurves over a range of frequencies.
The 3D framework allows us to explore how ECME is affected due to refraction in the stellar magnetospheric plasma for any arbitrary plasma density distribution. However, it makes a number of simplifying assumptions:
\begin{enumerate}
    \item The stellar magnetic field is an axi-symmetric dipole.
    \item The intrinsic ECME spectrum is flat.
    \item There is no frequency dependence of the intrinsic beaming angles, and the emission is beamed following the tangent plane beaming model \citep{trigilio2011}.
    \item The radiation travels in straight lines until it `hits' the closed stellar magnetosphere bounded by the magnetic field lines with equatorial radii equal to the Alfv\'en radius ($R_\mathrm{A}$).
\end{enumerate}
While the validity of the assumptions 2--4 is uncertain, assumption 1 is known to be invalid for the star HD\,142990, as the star's magnetic field is known to have a significant quadrupolar component \citep{shultz2018}. It is thus non-trivial to compare simulation and observation since non-dipolar geometry, as well as other factors such as the intrinsic spectrum and beaming pattern, are very likely to affect the final observed properties. We hence use the simulation as a means to gain insights about how different stellar parameters qualitatively affect the wideband properties of ECME solely due to propagation effects for an idealized scenario of axi-symmetric dipolar magnetic field.

The primary inputs to the framework are a density grid, a polar magnetic field strength ($B_\mathrm{p}$), an obliquity ($\beta$), Alfv\'en radius ($R_\mathrm{A}$), the equatorial radii of the magnetic field lines along which the auroral circles producing ECME are located ($L$), the magneto-ionic mode, the harmonic number, and the frequency of the observation. We assume that ECME is produced in the extraordinary mode at the second harmonic. $B_\mathrm{p}$, $\beta$ and $R_\mathrm{A}$ are set to 4.7 kG, $83^\circ$ and 22 $R_*$, respectively. To reduce computation time, we use a single value of $L$, set to $30\,R_*$. For the density grid, we use the `Rigidly Rotating Magnetosphere' \citep[RRM,][]{townsend2005} model to obtain the relative density distribution. Following \citet{das2024}, we use the following expression for the magnetospheric density:
\begin{align*}
    n_\mathrm{p}&=\frac{n_\mathrm{p0}}{r}(1+E\tilde{n}),
\end{align*}
where $\tilde{n}$ is the relative density predicted by the RRM model, $E$ is the density enhancement factor and $n_\mathrm{p0}$ is the overall density scaling factor. $r$ is the radial distance in units of stellar radius. While calculating $\tilde{n}$, we use a Kepler radius of $2.58\,R_*$ for HD\,142990.

By trial and error, the best agreement between simulated and observed lightcurves (in terms of the number of pulses observed at a given polarization and the relative location of the secondary pulses with respect to the primary pulses) are obtained for $n_\mathrm{p0}=10^9\,\mathrm{cm^{-3}}$, $E=10^4$ and by using an inclination angle of $i=36^\circ$ (top panel of Figure \ref{fig:best_case_sim_lc}). 
Note that, in the framework of \citet{das2020a}, phase zero corresponds to the positive extremum of \bz. However, for the convenience of comparing with observations, we shift the phase axis so that phase zero corresponds to the negative extremum of \bz~\citep[thus matches with the phase zero of the ephemeris used,][]{shultz2019_0}.
The bottom panel of Figure \ref{fig:best_case_sim_lc} shows the lightcurves using $i=55^\circ$, which is the value reported for HD\,142990 \citep{shultz2019c}. The lightcurves obtained for the case of $i=55^\circ$ are clearly significantly different from those observed (shown in dashed curves in Figure \ref{fig:best_case_sim_lc}).
This discrepancy could be related to our assumption of a purely dipolar magnetic field.

\begin{figure*}
    \centering
    \includegraphics[width=0.75\textwidth]{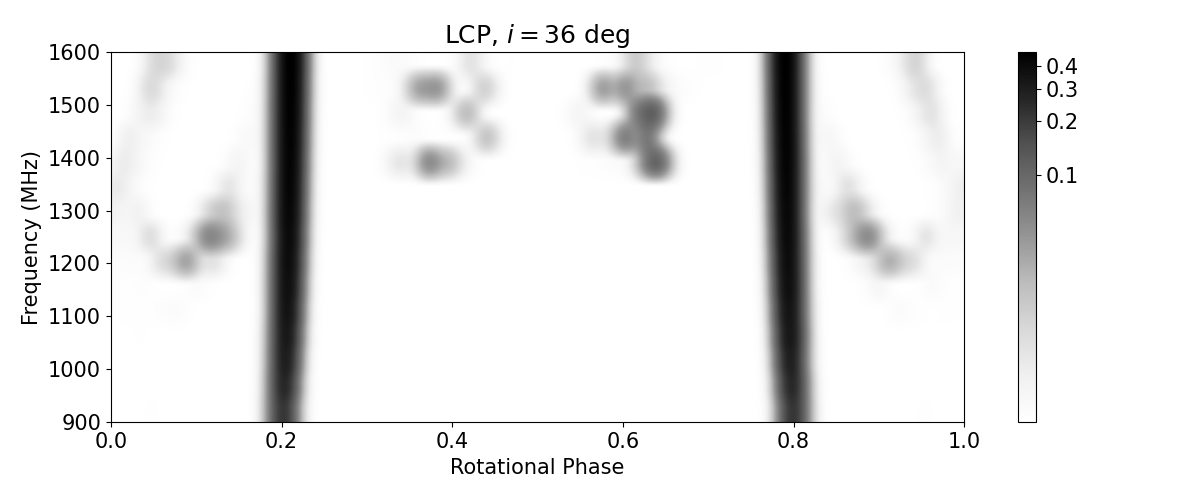}
    \includegraphics[width=0.75\textwidth]{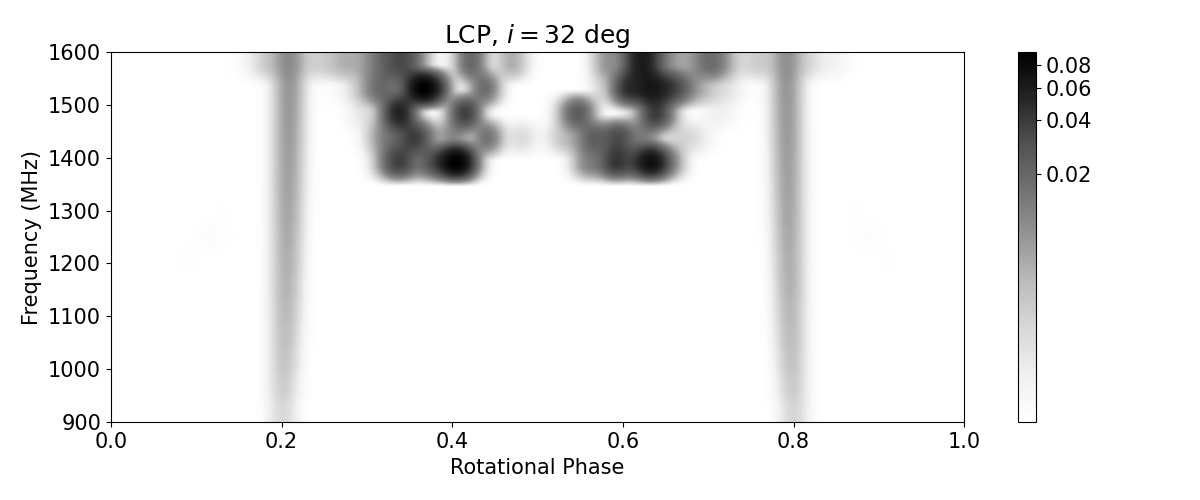}
    \includegraphics[width=0.75\textwidth]{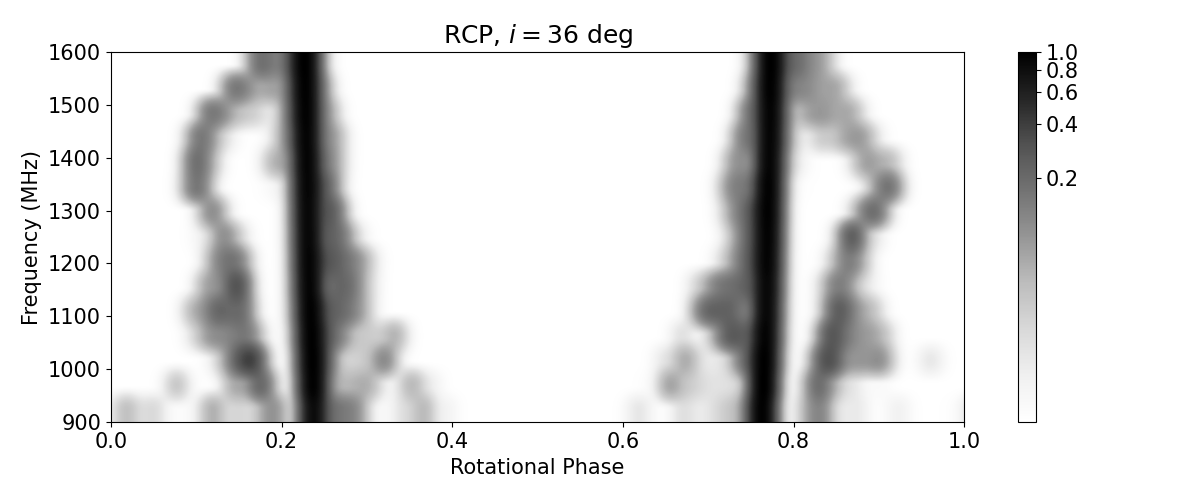}
    \includegraphics[width=0.75\textwidth]{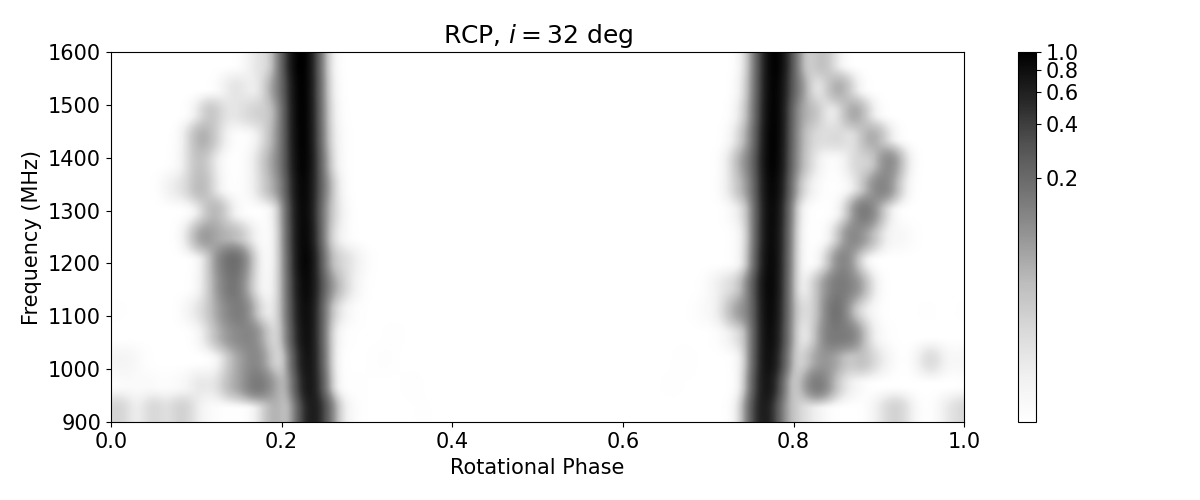}
    \caption{The simulated dynamic spectra for LCP and RCP emission from HD\,142990 for two different values of the inclination angle $i$. The details of the simulation is given in \S\ref{subsec:simulation}.}
    \label{fig:sim_DS}
\end{figure*}

The strong variation in the observed properties of ECME with different inclination angles can be seen more clearly from the simulated dynamic spectra shown in Figure \ref{fig:sim_DS}. By changing the inclination angle from $36^\circ$ (as used to obtain the lightcurves in Figure \ref{fig:best_case_sim_lc}) to $32^\circ$, following key changes occur:
\begin{enumerate}
    \item For the LCP emission, the primary pulses become significantly weaker for the lower value of the inclination angle. The secondary enhancements above 1400 MHz become much more prominent and those between 1200 and 1300 MHz disappear for $i=32^\circ$.
    \item For the RCP emission, lowering the inclination angle makes the primary pulses more prominent and weakens the secondary enhancements.
\end{enumerate}
Thus, the effects of changing the inclination angle on the LCP and RCP emission are qualitatively opposite. These simulated dynamic spectra correctly reproduces the direction of `sweep' in the frequency-time domain for the primary pulses, i.e. the LCP pulses are closely spaced at higher frequency as compared to that at lower frequencies (up to around 1300 MHz). They, however, differ from the observed dynamic spectra (Figure \ref{fig:DS}) in several aspects. For example, the observed dynamic spectrum for one of the LCP pulses shows that the direction of sweep of the primary pulse reverses at higher frequencies (top panel of Figure \ref{fig:LCP_DS_LC_day2}). This is not seen in the simulation for the primary pulse (top two rows of Figure \ref{fig:sim_DS}). It is, however, to be noted that for the case of $i=32^\circ$, the secondary pulses appear to sweep in the reverse direction w.r.t. the corresponding primary pulse. Thus, one possible explanation could be that the intrinsic spectrum of the primary LCP pulse is such that it is not visible above 1300 MHz and the `secondary' enhancements become the `primary' enhancements above that frequency.

The above example demonstrates the challenges of comparing simulation and observation due to the simplifying assumptions made in the former (e.g. flat intrinsic spectrum for ECME). Nevertheless, our simulation demonstrated that even for the most ideal situation, we can obtain a highly complicated dynamic spectra solely due to the highly azimuthally asymmetric magnetospheric density distribution. These effects will likely be present in HD\,142990 because of its large obliquity, but the true emission will be further complicated by several factors such as deviation of the magnetic field from dipolar geometry, intrinsic ECME spectrum, frequency dependent beaming angle, etc. 
In addition to that, emission at higher harmonics also remains a possibility \citep{das2021}. A more realistic simulation that takes into account the known non-dipolar geometry of the stellar magnetic field will also be useful to investigate such alternate possibilities as the cause behind the secondary enhancements.

\subsection{Origin of fine structures in ECME}\label{subsec:fine_structures}
Although fine structures in ECME from magnetic hot stars have not been reported before, such structures are known to exist in ECME observed from much cooler magnetic stars and planets. 
Fine structures with a timescales of a few seconds, bandwidths of $\lesssim$ kHZ and varying drift rates were reported for auroral kilometric radiation (AKR) by \citet{gurnett1979,gurnett1981}. Subsequent higher temporal and spectral resolution observations revealed fine structures at much smaller scales, which are associated with elementary AKR radiation sources \citep{pottelette1999,treumann2006}. Fine structures have also been reported for ECME observed from other solar system planets such as Jupiter \citep[e.g.][]{zarka1996} and late-type stars \citep[e.g.][etc.]{osten2008,callingham2021,bastian2022,zhang2023,bloot2024}. In particular, Figure 1 of \citet{zhang2023} demonstrates how increasing time resolution of observations reveals finer structures in the radio emission from the star AD\,Leo during a burst.

In view of the above, observation of fine structures in ECME from a magnetic hot star is not a surprise. Probably a more interesting observation is that the fine structures appear only for the LCP pulse near null 2 at frequencies close to its cut-off frequency. 
A possible scenario that explains this aspect is that there could be two emission components (both potentially of coherent nature), one responsible for the smooth emission and the other responsible for the spiky features. The former has a lower cut-off frequency and is stronger than the latter except for frequencies close to their cut-offs. Consequently, the fine structures become prominent only beyond a certain frequency that is close to the cut-off frequency of the smooth component.

An alternative scenario is that the smooth emission is actually composed of many finer spikes spaced closely to each other. Each such spike corresponds to individual emission sites, which are located along the different magnetic azimuths ($\phi$ coordinate in the magnetic frame with the dipole axis lying along the $Z$ axis) on a given auroral ring in the stellar magmetosphere. The location of the auroral rings and their radii are determined by several factors such as the frequency of emission, harmonic number, the stellar magnetic field strength, as well as the sites of particle acceleration. With increasing frequencies, the density of the elementary ECME sources on the auroral rings decreases. Thus, up to a certain frequency, the emission beams produced by nearby sites overlap so that we see a smooth emission. Beyond that frequency, the emission sites are too sparse, which is seen as fragmentation of the smooth component. A further decrease in the number of elementary ECME sources (as we go to even higher frequencies) eventually causes the cut-off. 

The above scenario of ECME with different intensities produced at different points on the auroral rings in the stellar magnetosphere is consistent with the idea of the emission being driven by small-scale magnetic reconnection triggered by centrifugal breakout (CBO). The CBO events are confined to small spatial scales and occur at different magnetic longitudes independent of each other \citep{shultz2020,owocki2020}.
Further insights regarding the origin of fine structures and the potential connection with cut-off frequencies can be obtained with higher time and spectral resolution wideband observations.
In the future, we aim to undertake detailed investigation of the fine structures so as to be able to establish their properties and eventually pinpoint their origin.

\section{Summary}\label{sec:summary}
HD\,142990 was one of the four MRPs, which was observed from sub-GHz to $\sim \,\mathrm{GHz}$ frequencies \citep{das2019a,das2023}. Although past studies involved observations only over rotational phase ranges surrounding the magnetic nulls, they already revealed a number of peculiar properties of the primary pulses, such as double-peaked pulse-profile \citep{das2019a}.
The non-ideal properties of the star was one of the key motivations for developing the 3D framework to investigate the effect of dense magnetospheric plasma on the ECME lightcurves \citep{das2020a}. 
This framework, in turn, predicted the existence of secondary enhancements for highly oblique rotators. This prediction provided the motivation to carry out full rotation cycle observation of MRPs to obtain the complete picture, instead of restricting the observations to only the magnetic nulls.

Though our primary aim was to search for secondary enhancements, our observations with a highly sensitive instrument like MeerKAT led to the serendipitous discovery of fine structures in the dynamic spectrum of one of the four primary pulses. Fine structures are known to be important characteristics of ECME, and have been reported from solar system planets and late-type stars. Before this study, magnetic hot stars were studied at time scales $\sim$ minutes. Thus, the dynamic spectra shown in Figure \ref{fig:DS} (time resolution of 8 seconds and spectral resolution of 418 kHz) represent the highest resolution dynamic spectra ever reported for ECME from magnetic hot stars. This capability was the primary reason for the discovery of fine structures. Note that the minimum width of the spikes is 8 seconds, which is also the integration time of our data. In the future, observation at even higher time and spectral resolutions should be conducted so as to investigate the presence of emission at smaller scales, and the cause behind the apparent absence of such structures for other pulses from the star.


Regarding the search for enhancements away from magnetic nulls, our observation showed that indeed the star also emits secondary radio pulses, which are visible at rotational phases significantly offset from the magnetic nulls (\S\ref{sec:results}).
Although these results are broadly consistent with the prediction of the 3D framework, we find a number of discrepancies between observed and simulated lightcurves across our observed frequency band (\S\ref{subsec:simulation}). Such discrepancies could arise due to the idealised assumptions made in the simulation. 
Nonetheless, the current simulation provided new insights about the crucial role of inclination angle, in addition to obliquity, in determining the rotational phases of arrival of ECME pulses. 

A complementary way to test the propagation effect scenario is to observe more MRPs with varying obliquities and inclination angles for their complete rotation cycles. A lower obliquity implies a more symmetric plasma distribution, reducing the importance of propagation effects. In the extreme case of aligned rotators (zero obliquity), the high density plasma is expected to be confined very close to the magnetic equatorial plane \citep{townsend2005}. As ECME is produced above the stellar magnetic poles, it is unlikely to encounter these high density regions in that case, and should propagate approximately in straight lines. In reality, aligned rotators will not produce the emission in the form of pulses \citep[e.g. the case of $\rho\,\mathrm{Oph C}$,][]{leto2020b}. Thus, to test the propagation effect scenario, we will need MRPs with non-zero obliquities only, but spanning a wide range over $\sim 10^\circ-90^\circ$. 
As mentioned already, the current simulation also shows the importance of considering the inclination angle while explaining the observed properties of ECME. 
Therefore, a thorough investigation will need a sample of MRPs also spanning a wide range of inclination angles, and possibly magnetic field strengths as well.

The ECME phenomenon is also observed from other types of stars and even planets. Magnetic hot stars offer us a stable environment to understand this phenomenon owing to their stable magnetic field. A complete understanding of the phenomenon, including the arrival phases of the pulses, will be important not only for the magnetic hot stars, but also for other types of systems. In particular, our work clearly showed that ECME can be observed at more than two sets of rotational phase ranges. The underlying reason(s) can have implications for other systems as well, such as in determining the arrival times of pulses arising due to star-planet interactions.

To summarize, we confirm that the visibility of ECME pulses from magnetic hot stars are not restricted to only the phases close to magnetic nulls. Our results also establish the need to conduct a much larger campaign to observe MRPs spanning a range of relevant stellar parameters (e.g. inclination angle, obliquity, magnetic field strength) for their complete rotation cycles, over wideband and at higher time and spectral resolutions, in order to fully understand the ECME phenomenon.

\begin{acknowledgements}
We thank the referee for their comments and suggestions that have helped us to significantly improve our manuscript. 
BD thanks Tim Bastian, Laura Driessen, Joshua Pritchard, Andrew Zic and Surajit Mondal for helpful discussions and suggestions.
The National Radio Astronomy Observatory is a facility of the National Science Foundation operated under cooperative agreement by Associated Universities, Inc.
The MeerKAT telescope is operated by the South African Radio Astronomy Observatory, which is a facility of the National Research Foundation, an agency of the Department of Science and Innovation.
This research has made use of NASA's Astrophysics Data System.
\end{acknowledgements}

\bibliography{das}

\begin{thebibliography}{}
\expandafter\ifx\csname natexlab\endcsname\relax\def\natexlab#1{#1}\fi
\providecommand{\url}[1]{\href{#1}{#1}}
\providecommand{\dodoi}[1]{doi:~\href{http://doi.org/#1}{\nolinkurl{#1}}}
\providecommand{\doeprint}[1]{\href{http://ascl.net/#1}{\nolinkurl{http://ascl.net/#1}}}
\providecommand{\doarXiv}[1]{\href{https://arxiv.org/abs/#1}{\nolinkurl{https://arxiv.org/abs/#1}}}

\bibitem[{{Bastian} {et~al.}(2022){Bastian}, {Cotton}, \& {Hallinan}}]{bastian2022}
{Bastian}, T.~S., {Cotton}, W.~D., \& {Hallinan}, G. 2022, \apj, 935, 99, \dodoi{10.3847/1538-4357/ac7d57}

\bibitem[{{Bloot} {et~al.}(2024){Bloot}, {Callingham}, {Vedantham}, {Kavanagh}, {Pope}, {Climent}, {Guirado}, {Pe{\~n}a-Mo{\~n}ino}, \& {P{\'e}rez-Torres}}]{bloot2024}
{Bloot}, S., {Callingham}, J.~R., {Vedantham}, H.~K., {et~al.} 2024, \aap, 682, A170, \dodoi{10.1051/0004-6361/202348065}

\bibitem[{{Callingham} {et~al.}(2021){Callingham}, {Pope}, {Feinstein}, {Vedantham}, {Shimwell}, {Zarka}, {Tasse}, {Lamy}, {Veken}, {Toet}, {Sabater}, {Best}, {van Weeren}, {R{\"o}ttgering}, \& {Ray}}]{callingham2021}
{Callingham}, J.~R., {Pope}, B.~J.~S., {Feinstein}, A.~D., {et~al.} 2021, \aap, 648, A13, \dodoi{10.1051/0004-6361/202039144}

\bibitem[{{Cotton}(2008)}]{OBIT}
{Cotton}, W.~D. 2008, PASP, 120, 439, \dodoi{10.1086/586754}

\bibitem[{Cotton {et~al.}(2020)Cotton, Thorat, Condon, Frank, Józsa, White, Deane, Oozeer, Atemkeng, Bester, Fanaroff, Kupa, Smirnov, Mauch, Krishnan, \& Camilo}]{XGalaxy}
Cotton, W.~D., Thorat, K., Condon, J.~J., {et~al.} 2020, MNRAS, 495, 1271, \dodoi{10.1093/mnras/staa1240}

\bibitem[{{Das} \& {Chandra}(2021)}]{das2021}
{Das}, B., \& {Chandra}, P. 2021, \apj, 921, 9, \dodoi{10.3847/1538-4357/ac1075}

\bibitem[{{Das} \& {Chandra}(2023)}]{das2023}
---. 2023, \apj, 957, 53, \dodoi{10.3847/1538-4357/acf929}

\bibitem[{{Das} {et~al.}(2024){Das}, {Chandra}, \& {Petit}}]{das2024}
{Das}, B., {Chandra}, P., \& {Petit}, V. 2024, \apj, 974, 267, \dodoi{10.3847/1538-4357/ad71c5}

\bibitem[{{Das} {et~al.}(2019){Das}, {Chandra}, {Shultz}, \& {Wade}}]{das2019a}
{Das}, B., {Chandra}, P., {Shultz}, M.~E., \& {Wade}, G.~A. 2019, \apj, 877, 123, \dodoi{10.3847/1538-4357/ab1b12}

\bibitem[{{Das} {et~al.}(2018){Das}, {Chandra}, \& {Wade}}]{das2018}
{Das}, B., {Chandra}, P., \& {Wade}, G.~A. 2018, \mnras, 474, L61, \dodoi{10.1093/mnrasl/slx193}

\bibitem[{{Das} {et~al.}(2020){Das}, {Mondal}, \& {Chandra}}]{das2020a}
{Das}, B., {Mondal}, S., \& {Chandra}, P. 2020, \apj, 900, 156, \dodoi{10.3847/1538-4357/aba8fd}

\bibitem[{{Das} {et~al.}(2022){Das}, {Chandra}, {Shultz}, {Wade}, {Sikora}, {Kochukhov}, {Neiner}, {Oksala}, \& {Alecian}}]{das2022}
{Das}, B., {Chandra}, P., {Shultz}, M.~E., {et~al.} 2022, \apj, 925, 125, \dodoi{10.3847/1538-4357/ac2576}

\bibitem[{{Gurnett} \& {Anderson}(1981)}]{gurnett1981}
{Gurnett}, D.~A., \& {Anderson}, R.~R. 1981, Geophysical Monograph Series, 25, 341, \dodoi{10.1029/GM025p0341}

\bibitem[{{Gurnett} {et~al.}(1979){Gurnett}, {Anderson}, {Scarf}, {Fredricks}, \& {Smith}}]{gurnett1979}
{Gurnett}, D.~A., {Anderson}, R.~R., {Scarf}, F.~L., {Fredricks}, R.~W., \& {Smith}, E.~J. 1979, \ssr, 23, 103, \dodoi{10.1007/BF00174114}

\bibitem[{{Jonas} \& {MeerKAT Team}(2016)}]{jonas2016}
{Jonas}, J., \& {MeerKAT Team}. 2016, in MeerKAT Science: On the Pathway to the SKA, 1, \dodoi{10.22323/1.277.0001}

\bibitem[{{Leto} {et~al.}(2016){Leto}, {Trigilio}, {Buemi}, {Umana}, {Ingallinera}, \& {Cerrigone}}]{leto2016}
{Leto}, P., {Trigilio}, C., {Buemi}, C.~S., {et~al.} 2016, \mnras, 459, 1159, \dodoi{10.1093/mnras/stw639}

\bibitem[{{Leto} {et~al.}(2017){Leto}, {Trigilio}, {Oskinova}, {Ignace}, {Buemi}, {Umana}, {Ingallinera}, {Todt}, \& {Leone}}]{leto2017}
{Leto}, P., {Trigilio}, C., {Oskinova}, L., {et~al.} 2017, \mnras, 467, 2820, \dodoi{10.1093/mnras/stx267}

\bibitem[{{Leto} {et~al.}(2018){Leto}, {Trigilio}, {Oskinova}, {Ignace}, {Buemi}, {Umana}, {Ingallinera}, {Leone}, {Phillips}, {Agliozzo}, {Todt}, \& {Cerrigone}}]{leto2018}
{Leto}, P., {Trigilio}, C., {Oskinova}, L.~M., {et~al.} 2018, \mnras, 476, 562, \dodoi{10.1093/mnras/sty244}

\bibitem[{{Leto} {et~al.}(2019){Leto}, {Trigilio}, {Oskinova}, {Ignace}, {Buemi}, {Umana}, {Cavallaro}, {Ingallinera}, {Bufano}, {Phillips}, {Agliozzo}, {Cerrigone}, {Todt}, {Riggi}, \& {Leone}}]{leto2019}
---. 2019, \mnras, 482, L4, \dodoi{10.1093/mnrasl/sly179}

\bibitem[{{Leto} {et~al.}(2020{\natexlab{a}}){Leto}, {Trigilio}, {Leone}, {Pillitteri}, {Buemi}, {Fossati}, {Cavallaro}, {Oskinova}, {Ignace}, {Krti{\v{c}}ka}, {Umana}, {Catanzaro}, {Ingallinera}, {Bufano}, {Agliozzo}, {Phillips}, {Cerrigone}, {Riggi}, {Loru}, {Munari}, {Gangi}, {Giarrusso}, \& {Robrade}}]{leto2020}
{Leto}, P., {Trigilio}, C., {Leone}, F., {et~al.} 2020{\natexlab{a}}, \mnras, 493, 4657, \dodoi{10.1093/mnras/staa587}

\bibitem[{{Leto} {et~al.}(2020{\natexlab{b}}){Leto}, {Trigilio}, {Buemi}, {Leone}, {Pillitteri}, {Fossati}, {Cavallaro}, {Oskinova}, {Ignace}, {Krti{\v{c}}ka}, {Umana}, {Catanzaro}, {Ingallinera}, {Bufano}, {Riggi}, {Cerrigone}, {Loru}, {Schillir{\'o}}, {Agliozzo}, {Phillips}, {Giarrusso}, \& {Robrade}}]{leto2020b}
{Leto}, P., {Trigilio}, C., {Buemi}, C.~S., {et~al.} 2020{\natexlab{b}}, \mnras, 499, L72, \dodoi{10.1093/mnrasl/slaa157}

\bibitem[{{Lim} {et~al.}(1996){Lim}, {Drake}, \& {Linsky}}]{lim1996}
{Lim}, J., {Drake}, S.~A., \& {Linsky}, J.~L. 1996, Astronomical Society of the Pacific Conference Series, Vol.~93, {Rotational Modulation of Radio Emission from the Magnetic BP Star HR 5624}, ed. A.~R. {Taylor} \& J.~M. {Paredes}, 324

\bibitem[{{Mauch} {et~al.}(2020){Mauch}, {Cotton}, {Condon}, {Matthews}, {Abbott}, {Adam}, {Aldera}, {Asad}, {Bauermeister}, {Bennett}, {Bester}, {Botha}, {Brederode}, {Brits}, {Buchner}, {Burger}, {Camilo}, {Chalmers}, {Cheetham}, {de Villiers}, {de Villiers}, {Dikgale-Mahlakoana}, {du Toit}, {Esterhuyse}, {Fadana}, {Fanaroff}, {Fataar}, {February}, {Frank}, {Gamatham}, {Geyer}, {Goedhart}, {Gounden}, {Gumede}, {Heywood}, {Hlakola}, {Horrell}, {Hugo}, {Isaacson}, {J{\'o}zsa}, {Jonas}, {Julie}, {Kapp}, {Kasper}, {Kenyon}, {Kotz{\'e}}, {Kriek}, {Kriel}, {Kusel}, {Lehmensiek}, {Loots}, {Lord}, {Lunsky}, {Madisa}, {Magnus}, {Main}, {Malan}, {Manley}, {Marais}, {Martens}, {Merry}, {Millenaar}, {Mnyandu}, {Moeng}, {Mokone}, {Monama}, {Mphego}, {New}, {Ngcebetsha}, {Ngoasheng}, {Ockards}, {Oozeer}, {Otto}, {Patel}, {Peens-Hough}, {Perkins}, {Ramaila}, {Ramudzuli}, {Renil}, {Richter}, {Robyntjies}, {Salie}, {Schollar}, {Schwardt}, {Serylak}, {Siebrits}, {Sirothia}, {Smirnov}, {Sofeya}, {Stone}, {Taljaard}, {Tasse},
  {Theron}, {Tiplady}, {Toruvanda}, {Twum}, {van Balla}, {van der Byl}, {van der Merwe}, {Van Tonder}, {Wallace}, {Welz}, {Williams}, \& {Xaia}}]{DEEP2}
{Mauch}, T., {Cotton}, W.~D., {Condon}, J.~J., {et~al.} 2020, ApJ, 888, 61, \dodoi{10.3847/1538-4357/ab5d2d}

\bibitem[{{McMullin} {et~al.}(2007){McMullin}, {Waters}, {Schiebel}, {Young}, \& {Golap}}]{mcmullin2007}
{McMullin}, J.~P., {Waters}, B., {Schiebel}, D., {Young}, W., \& {Golap}, K. 2007, in Astronomical Society of the Pacific Conference Series, Vol. 376, Astronomical Data Analysis Software and Systems XVI, ed. R.~A. {Shaw}, F.~{Hill}, \& D.~J. {Bell}, 127

\bibitem[{{Osten} \& {Bastian}(2008)}]{osten2008}
{Osten}, R.~A., \& {Bastian}, T.~S. 2008, \apj, 674, 1078, \dodoi{10.1086/525013}

\bibitem[{{Owocki} {et~al.}(2020){Owocki}, {Shultz}, {ud-Doula}, {Sundqvist}, {Townsend}, \& {Cranmer}}]{owocki2020}
{Owocki}, S.~P., {Shultz}, M.~E., {ud-Doula}, A., {et~al.} 2020, \mnras, 499, 5366, \dodoi{10.1093/mnras/staa2325}

\bibitem[{{Pottelette} {et~al.}(1999){Pottelette}, {Ergun}, {Treumann}, {Berthomier}, {Carlson}, {McFadden}, \& {Roth}}]{pottelette1999}
{Pottelette}, R., {Ergun}, R.~E., {Treumann}, R.~A., {et~al.} 1999, \grl, 26, 2629, \dodoi{10.1029/1999GL900462}

\bibitem[{{Reynolds}(1994)}]{Reynolds94}
{Reynolds}, J.~E. 1994, {ATNF Memo}, {AT/39.3/040}

\bibitem[{{Shultz} {et~al.}(2019{\natexlab{a}}){Shultz}, {Rivinius}, {Das}, {Wade}, \& {Chand ra}}]{shultz2019_0}
{Shultz}, M., {Rivinius}, T., {Das}, B., {Wade}, G.~A., \& {Chand ra}, P. 2019{\natexlab{a}}, \mnras, 486, 5558, \dodoi{10.1093/mnras/stz1129}

\bibitem[{{Shultz} {et~al.}(2018){Shultz}, {Wade}, {Rivinius}, {Neiner}, {Alecian}, {Bohlender}, {Monin}, {Sikora}, {MiMeS Collaboration}, \& {BinaMIcS Collaboration}}]{shultz2018}
{Shultz}, M.~E., {Wade}, G.~A., {Rivinius}, T., {et~al.} 2018, \mnras, 475, 5144, \dodoi{10.1093/mnras/sty103}

\bibitem[{{Shultz} {et~al.}(2019{\natexlab{b}}){Shultz}, {Wade}, {Rivinius}, {Alecian}, {Neiner}, {Petit}, {Owocki}, {ud-Doula}, {Kochukhov}, {Bohlender}, {Keszthelyi}, {MiMeS Collaboration}, \& {BinaMIcS Collaboration}}]{shultz2019c}
---. 2019{\natexlab{b}}, \mnras, 490, 274, \dodoi{10.1093/mnras/stz2551}

\bibitem[{{Shultz} {et~al.}(2020){Shultz}, {Owocki}, {Rivinius}, {Wade}, {Neiner}, {Alecian}, {Kochukhov}, {Bohlender}, {ud-Doula}, {Landstreet}, {Sikora}, {David-Uraz}, {Petit}, {Cerraho{\u{g}}lu}, {Fine}, {Henson}, {MiMeS Collaboration}, \& {BinaMIcS Collaboration}}]{shultz2020}
{Shultz}, M.~E., {Owocki}, S., {Rivinius}, T., {et~al.} 2020, \mnras, 499, 5379, \dodoi{10.1093/mnras/staa3102}

\bibitem[{{Townsend} \& {Owocki}(2005)}]{townsend2005}
{Townsend}, R.~H.~D., \& {Owocki}, S.~P. 2005, \mnras, 357, 251, \dodoi{10.1111/j.1365-2966.2005.08642.x}

\bibitem[{{Treumann}(2006)}]{treumann2006}
{Treumann}, R.~A. 2006, \aapr, 13, 229, \dodoi{10.1007/s00159-006-0001-y}

\bibitem[{{Trigilio} {et~al.}(2000){Trigilio}, {Leto}, {Leone}, {Umana}, \& {Buemi}}]{trigilio2000}
{Trigilio}, C., {Leto}, P., {Leone}, F., {Umana}, G., \& {Buemi}, C. 2000, \aap, 362, 281.
\newblock \doarXiv{astro-ph/0007097}

\bibitem[{{Trigilio} {et~al.}(2011){Trigilio}, {Leto}, {Umana}, {Buemi}, \& {Leone}}]{trigilio2011}
{Trigilio}, C., {Leto}, P., {Umana}, G., {Buemi}, C.~S., \& {Leone}, F. 2011, \apjl, 739, L10, \dodoi{10.1088/2041-8205/739/1/L10}

\bibitem[{{Villadsen} \& {Hallinan}(2019)}]{villadsen2019}
{Villadsen}, J., \& {Hallinan}, G. 2019, \apj, 871, 214, \dodoi{10.3847/1538-4357/aaf88e}

\bibitem[{{Zarka} {et~al.}(1996){Zarka}, {Farges}, {Ryabov}, {Abada-Simon}, \& {Denis}}]{zarka1996}
{Zarka}, P., {Farges}, T., {Ryabov}, B.~P., {Abada-Simon}, M., \& {Denis}, L. 1996, \grl, 23, 125, \dodoi{10.1029/95GL03780}

\bibitem[{{Zhang} {et~al.}(2023){Zhang}, {Tian}, {Zarka}, {Louis}, {Lu}, {Gao}, {Sun}, {Yu}, {Chen}, {Cheng}, \& {Wang}}]{zhang2023}
{Zhang}, J., {Tian}, H., {Zarka}, P., {et~al.} 2023, \apj, 953, 65, \dodoi{10.3847/1538-4357/acdb77}

\end{thebibliography}
\bibliographystyle{aasjournal}


\end{document}